# Variable domain N-linked glycosylation and negative surface charge are key features of monoclonal ACPA: implications for B-cell selection

Katy A. Lloyd[1], Johanna Steen[1], Khaled Amara[1], Philip J. Titcombe[1,2], Lena Israelsson[1], Susanna L. Lundström[3], Diana Zhou[1], Roman A. Zubarev[3], Evan Reed[1], Luca Piccoli[4], Cem Gabay[5], Antonio Lanzavecchia[4], Dominique Baeten[6], Karin Lundberg[1], Daniel L. Mueller[2], Lars Klareskog[1], Vivianne Malmström[1], and Caroline Grönwall[1*]

*1. Department of Medicine, Rheumatology Unit, Center for Molecular, Medicine Karolinska Institutet, Karolinska University Hospital, Stockholm, Sweden*
*2. The Center for Immunology, University of Minnesota Medical School, Minneapolis, Minnesota, USA*
*3. Department of Medical Biochemistry and Biophysics, Karolinska Institutet, Stockholm, Sweden*
*4. Institute for Research in Biomedicine, Università della Svizzera italiana, Bellinzona, Switzerland*
*5. Division of Rheumatology, University Hospitals of Geneva, Geneva, Switzerland*
*6. Department of Clinical Immunology and Rheumatology, Academic Medical Center, University of Amsterdam, Amsterdam, The Netherlands.*

*** Corresponding author:** Caroline Grönwall, *Department of Medicine, Rheumatology Unit, Karolinska Institutet and Karolinska University Hospital, Center for Molecular Medicine L8:04, 17176 Stockholm, Sweden.* caroline.gronwall@ki.se

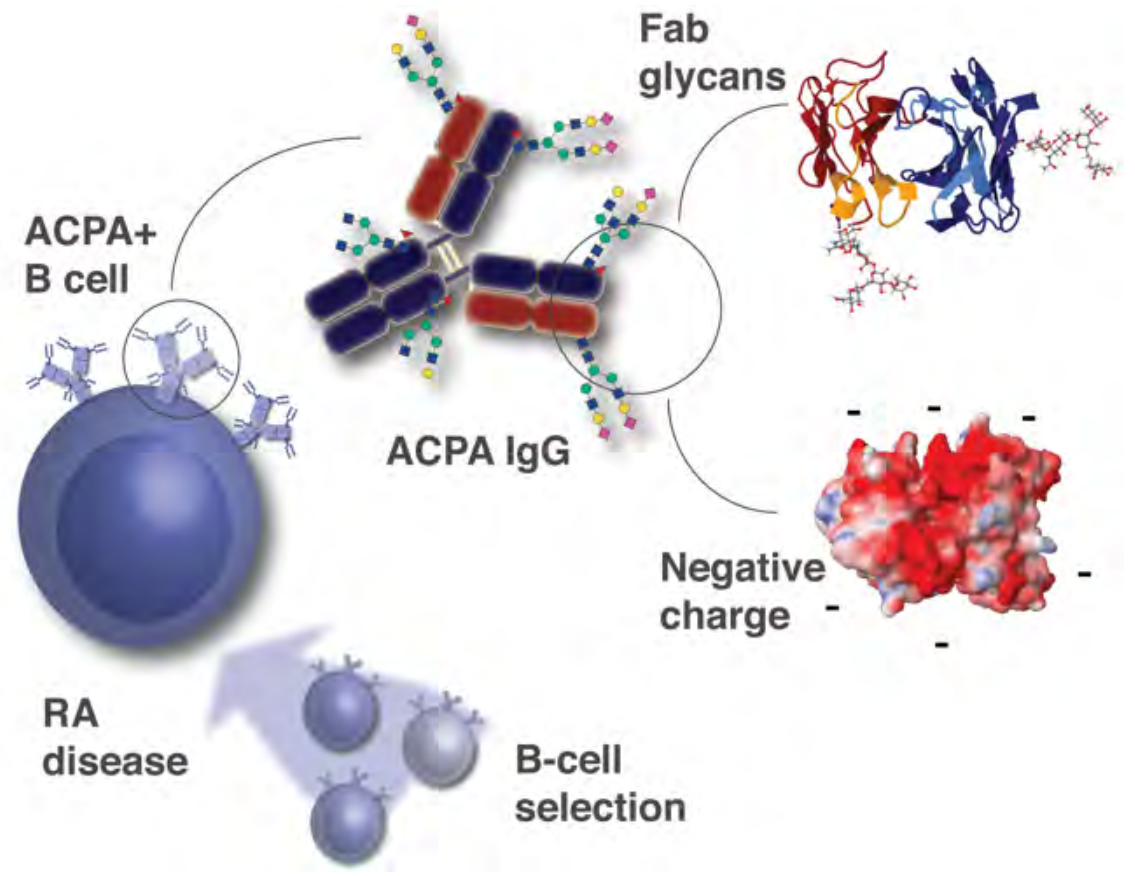

***Anti-citrullinated protein autoantibodies (ACPA) in rheumatoid arthritis have unique properties including negative surface charge and sialic acid-containing Fab N-glycosylation that does not influence CCP-binding. These features are introduced by somatic hypermutations, yet are not displayed by other highly mutated antibodies. Consequently, specific selection pressure(s) must drive ACPA+ B cells.***

Autoreactive B cells have a central role in the pathogenesis of rheumatoid arthritis (RA), and recent findings have proposed that anti-citrullinated protein autoantibodies (ACPA) may be directly pathogenic. Herein, we demonstrate the frequency of variable-region glycosylation in single-cell cloned mAbs. A total of 14 ACPA mAbs were evaluated for predicted N-linked glycosylation motifs *in silico* and compared to 452 highly-mutated mAbs from RA patients and controls. Variable region N-linked motifs (N-X-S/T) were strikingly prevalent within ACPA (100%) compared to somatically hypermutated (SHM) RA bone marrow plasma cells (21%), and synovial plasma cells from seropositive (39%) and seronegative RA (7%). When normalized for SHM, ACPA still had significantly higher frequency of N-linked motifs compared to all studied mAbs including highly-mutated HIV broadly-neutralizing and malaria-associated mAbs. The Fab glycans of ACPA-mAbs were highly sialylated, contributed to altered charge, but did not influence antigen binding. The analysis revealed evidence of unusual B-cell selection pressure and SHM-mediated decreased in surface charge and isoelectric point in ACPA. It is still unknown how these distinct features of anti-citrulline immunity may have an impact on pathogenesis. However, it is evident that they offer selective advantages for ACPA+ B cells, possibly also through non-antigen driven mechanisms.

**Keywords:** Fab glycosylation, N-linked glycosylation, anti-citrullinated protein autoantibodies, anti-CCP, Autoreactive B cells

## Introduction

Rheumatoid arthritis (RA) is a destructive inflammatory disease affecting 0.5-1% of the population and characterized by synovial inflammation and joint destruction (reviewed in [1]). Autoantibodies to proteins post-translationally modified by citrullination, denoted anti-citrullinated protein antibodies (ACPA), are found in ~65-80% of established RA patients and are included in the current ACR/EULAR RA classification criteria [2]. ACPA seropositive disease has a distinct pathogenesis compared to seronegative RA, and is associate with increased joint destruction and poorer prognosis (reviewed in [3]). Although still controversial, ACPA have been postulated to be directly involved in seropositive RA pathogenesis, and some ACPA can modulate disease by enhancing bone destruction and inducing pain in model systems [4, 5]. The broadening of ACPA repertoires prior to clinical onset of RA [6],and the relative success of B cell targeted treatment in certain seropositive RA patients [7], may further highlight a role for autoantibodies and B cells in RA.

The ability of IgG Fc-glycans to modulate antibody effector functions is well characterized. Variations in the Fc-glycan profiles can augment inflammation and contribute to pathogenesis in autoimmune disease. Indeed, polyclonal ACPA-IgG exhibit decreased terminal galactose and sialic acid residues when compared to total serum IgG, consistent with a pro-inflammatory profile [8, 9].
In contrast to Fc-glycosylation that is present in some form in all IgG, Fab glycosylation is only found in ~15-25% of serum IgG (reviewed in [10]). Although the consequence of Fab glycosylation remains undetermined, it may alter IgG functionality by modulating antigen binding or by extending antibody half-life [11-14]. Previous studies have documented that N-linked glycan motifs (N-glyc) can be present on both heavy and light chains [15, 16]. The majority of variable region N-glyc sites are thought to arise through somatic hypermutation (SHM), as only a small number of variable region germline alleles encode for amino acid sequences featuring N-glyc sites [17].
Monoclonal ACPA-IgG exhibit several unique features including a remarkable cross-reactivity between different citrullinated proteins and peptides as well as a high rate of SHM. Herein, we further explore the biochemical characteristics of ACPA antibodies directly linked to the high level of somatic mutations. We report on the Fab glycosylation frequency and isoelectric charge properties of monoclonal ACPA compared to other single-cell derived antibodies, characterize the ACPA glycan-profile, and discuss potential immunobiological implications.

## Results

### ACPA-IgG exhibit extensive N-linked glycan motifs

To evaluate the frequency of N-glycosylation sites in different B cell populations we screened sequences from a range of expressed monoclonal antibodies derived from the blood and synovium of RA patients and which previously have been isolated from single B cells. In addition, the analysis included single cell BCR sequences from RA bone marrow plasma cells, and blood memory B cells from a healthy individual. For comparison, we utilized paired Ig sequences from the literature for antibodies with known specificities from HIV patients and malaria inflected patients [18, 19]. Some RA antibodies have been published elsewhere but this study also included a number of previously unpublished clones. The investigation focused on 14 identified ACPA mAbs isolated from five different CCP-positive RA patients. Four of the ACPA clones were derived from synovial plasma cells (PC), eight from peripheral antigen-tetramer sorted memory B cells, and two from memory B cells (memB; Table 1). All ACPA were specific for citrullinated peptides and proteins by ELISA or antigen microarray assay and did not bind to arginine peptide versions, had no rheumatoid factor activity, or showed any unspecific polyreactivity to other control antigens. However, they were all cross-reactive to multiple citrullinated antigens.

Consistent with extensive SHM, ACPA clones displayed on average 52 mismatches compared to the closest predicted germline heavy chain rearrangement and 34 mismatches in the light chain sequences (Table 2). The N-glyc frequencies in ACPA were compared to single cell-derived mAbs from different B cells groups and compartments from RA patients and controls (Table 2, Supporting Information Table 1). These were pre-selected for higher mutation rates (>15 SHM in the VH or VL) for better comparisons with the highly-mutated ACPA. Similarly, the analyzed sequences from broadly neutralizing mAbs from HIV patients (HIV bnAb) and the *Plasmodium falciparum* specific memory B cell sequences from malaria patients (Pf memB) were all highly mutated [18, 19].
The consensus N-glyc motif (N-*X*-S/T) was observed in the variable region of all 14 ACPA (100%), whereby non-ACPA synovial memory B cells (syn. memB, 15% n=20) and bone marrow plasma cell (BM PC, 21% n=198) sequences derived from CCP+ RA patients, malaria-derived memory B cell sequences (Pf memB, 25%, n=103), and healthy memory B cell sequences (19%, n=27) were within the 15-25% reference range for Fab glycosylation (Table 2, Fig. 1A) [20, 21]. Intriguingly, circulating memory B cells (memB, 32%, n=25) and synovial plasma cells from CCP+ RA patients (syn. PC, 39% n=33), as well as HIV bnAb sequences (63% n=19), contained increased rates of variable region N-glyc compared to circulating memory B cells in healthy individuals, but still lower than that of the ACPA (Table 2, Fig. 1A). When adjusting for SHM rates (Table 1, Fig. 1A), ACPA exhibited significantly higher normalized N-glyc rates in paired V sequences compared to CCP+ synovial PC (p<0.0001), HIV bnAb (p=0.001), CCP+ blood memB (p<0.0001), synovial memB, BM PC, Pf memB, and healthy memB (p<0.0001, respectively). An increased frequency of N-glyc in ACPA was observed also when analyzing VH or VL regions separately (Fig 1B, C). N-glyc in ACPA were all acquired through SHM (Table 1). Similarly, among the HIV bnAbs most N-glyc were also SHM-derived, but one HIV bnAb clone had a N-glyc that was featured in the germline sequence (Supporting Information Table 2). For the HIV bnAb, the light chain sequences contained a significantly increase in SHM-normalized N-glyc numbers compared to other memB sequences (p<0.0002, Supporting Information Fig. 1B).
HIV bnAbs displayed higher N-glyc frequencies in the light chain (42%) compared to the heavy chain (32%, Fig 1B, C), while the ACPA had equal frequencies in the heavy chain and light chain sequences (71%, Fig 1B, C). Notably, some ACPA mAbs contained up to three N-glyc motifs per chain (Fig. 1D). HIV bnAb also featured multiple N-glyc per chain, albeit at a lower rate than the ACPA. There was no striking difference in the location of N-glyc within variable region sequences (Fig. 1E), with ACPA N-glyc featured in both FWRs and CDRs. All verified ACPA clones had glycosylations sites and there was consequently no difference in the frequency of N-glyc sites in ACPA derived from synovium or blood B cells. Similarly, no significant difference in SHM-normalized N-glyc rates was observed for ACPA derived from synovial plasma cells compared to those derived from blood memory B cells (Supporting Information Fig. 2A). Furthermore, the significant differences between ACPA and control mAbs remained consistent when analyzing blood memory B cells or synovial plasma cells separately towards the ACPAs derived from the same compartment (Supporting Information Fig. 2 B and C). Intriguingly, synovial PC mAb isolated from seronegative patients exhibited a remarkably low frequency on N-glyc sites not only compared to ACPA (Table 2), but also towards synovial non-ACPA PC mAbs isolated from seropositive individuals (p=0.046, Supporting Information Fig. 2B). This suggest that although Fab glycosylation does not vary significantly between different RA B cell compartments, having a seropositive RA status has an impact.
Notably, as this study focused on analyzing N-glyc site frequencies in the translated variable region amino acid sequences from single cells sequencing, there may consequently be differences in the occupancy and composition of the N-glyc sites in the original BCRs on the B cells.

**Fab glycans of recombinant monoclonal ACPA-IgG are highly sialylated**

The presence of N-glyc motifs within sequences does not guarantee that these sites are occupied with glycans. To confirm whether the variable region N-glyc motifs in the ACPA were accessible for glycosylation, recombinant ACPA-IgG were analyzed for presence of Fab glycans. Expressed ACPA-IgG with predicted Fab N-glyc exhibited increased mass compared to the control mAb on SDS-PAGE (Fig. 2A; 1362:01E02). A number of bands with different molecular weight were observed for certain ACPA, potentially signifying the presence of different glycoforms of these mAbs. Furthermore, PNGase F digestion (removing all N-glycans) resulted in a mobility shift of ACPA-IgG gamma chains to a comparable level to the control mAb, whereas Endo S digestion (removing only Fc-glycans) produced partial reductions in IgG gamma chain molecular weight (Fig. 2A). This is a reliable and well-established method for analyzing Fc-glycans compared to Fab-glycans without the need for generation of (Fab')2 fragments. However, one limitation was that to ensure that functionality of the mAbs were maintained, the PNGase treatment was performed under native conditions instead of denaturing conditions, which could lead to only a partial enzyme digestion for some clones where the sites are protected by the structure. Mobility shifts in the light chain with PNGase treatment could also be observed for certain clones with predicted N-glyc sites in the VL sequence.

Lectin assays and mass spectrometry were utilized to analyze ACPA Fab glycan composition. SNA-ELISA suggested that the recombinant ACPA featured a high degree of sialylation, in contrast to control mAbs (Fig. 2B). SNA immunoblotting showed that certain ACPA mAbs contain sialylated residues on either the light or heavy chain (Fig. 2C), which corresponded with the presence of predicted N-glyc motifs (Fig. 2, Table 1). Importantly, non-ACPA control mAbs without predicted N-glyc sites in their variable region which were expressed in the same cell system gave no signal in SNA-ELISA or SNA-immunoblots, consistent with the low sialic acid content normally seen in recombinant IgG1 Fc-glycans. Curiously, certain ACPA-IgG expression batches had lower SNA reactivity but still with high ConA mannose-reactivity compared to control IgG and a mobility shift between EndoS and PNGase treatment (Fig. 2A, B, Supporting Information Fig. 3), suggesting variations in the Fab glycan composition. Notably, all ACPA batches displayed some level of variable region glycosylation (Supporting Information Fig. 4). Together, these data demonstrate that all of the screened recombinant ACPA-IgG had increased size and/or lectin-reactivity consistent with occupied variable region glycosylation sites (summarized Table 1). To validate, we analyzed CCP2 affinity purified polyclonal ACPA-IgG from pooled RA patients using the same lectin assays, and this also exhibited higher levels of sialic acid content compared to the control IgG flow-through from the CCP2 column (Supporting Information Fig. 5).

A sensitive glycopeptide profiling liquid chromatography mass spectrometry approach was used to characterize both the Fc and Fab glycan profiles of ACPA mAbs which confirmed the high sialic acid content of Fab glycans (Table 3, Supporting Information Table 3, Supporting Information Fig. 6). Additionally, ACPA Fab glycan compositions were observed to be distinctly different from Fc glycans. While the Fc glycans remained consistent between mAb clones and different recombinant production batches, a greater variation was detected in the Fab glycan composition. Mass spectrometry further revealed large intra-molecular variation with different composition at different N-glyc sites for clones with multiple sites (Supporting Information Table 3). Notably, all expressed monoclonal IgG used in the current study were expressed at the Karolinska Institute research laboratory, using the same cell system and a standardized protocol. The Fab glycopeptide profile of recombinantly expressed IgG from HEK2 cell systems cannot be expected to have the exact same glycan-composition as what has been reported for PNGase F released Fab-glycans analyzed from human polyclonal ACPA-IgG [22]. Yet, many of the complex sialylated forms identified herein have been detected in RA patients [23]. In any case and more importantly, our results clearly indicate that the monoclonal IgG indeed have occupied high-sialic acid Fab-glycosylation sites.

**ACPA-IgG glycosylation does not influence antigen binding**

We speculated that the introduction of bulky glycans into ACPA variable regions could either contribute to the binding interface, or block interactions with antigens by steric hindrance. All of the investigated ACPA except the clone 1003:10A01 have high anti-CCP3 reactivity (Fig. 3B). However, our analysis shows that deglycosylation of ACPA-IgG with PNGase F appeared to have no significant effect on CCP3-binding. Similar results were obtained in other citrulline binding assays (data not shown). One clone (62CFCT01E04) had an increased trend in binding following removal of glycans, yet this mAb appears to be an exception (Fig. 3B).

To further investigate why Fab glycosylation may not influence antigen binding, glycan modelling was added to homology-based structures of the ACPA based on the sequence data. Whilst we cannot fully depict the structure of the antigen-binding surface without a resolved structure, it is well established that the antigen Ig-binding surface is generally built up by heavy and light chain CDR-loops. Therefore, surfaces further away from the loops are less likely to be directly involved in antigen recognition. Notably, the models suggested that N-glyc were positioned primarily outside of the predicted antigen-binding regions, offering a potential explanation as to why ACPA mAb deglycosylation did not significantly influence antigen binding (Fig. 3A). N-glyc were positioned closer to the predicted antigen binding region for some ACPA (1325:04C03, 37CEPT02C04, 37CEPT01G09, 1003:10A01), yet these still do not appear to influence antigen binding.

**Effects of Fab glycosylation on B-cell selection pressures**

Since all ACPA N-glyc were introduced through SHM, the presence of variable region glycans may offer a selective advantage to ACPA+ B cells. We utilized the BASELINe tool [24] to investigate the distribution patterns of replacement/silent mutations as a measurement of selection pressure, comparing ACPA sequences to non-ACPA RA-derived mAbs, using methodology previously described [25].
Antigen-driven selection processes typically generate a higher level of replacement mutations compared to silent mutations in the CDRs, resulting in positive and negative selection strengths within CDRs and FWRs, respectively [24]. BASEline uses an algorithm that provides a statistical quantification of these somatic hypermutation patterns and presents the results as probability distribution graphs of selection strengths that can be compared between immunoglobulin groups. A CDR graph more to the right will indicate stronger positive selection in the CDRs and FRW graph more to the left will indicate a stronger negative selection pressure in the FRW.
Intriguingly, despite the high mutation levels, ACPA sequences featured negative selection strengths in the CDRs, especially when analyzing VH sequences (Fig. 4A), and the negative selection strengths were significantly different in ACPA versus non-ACPA RA sequences. When comparing ACPA to non-ACPA from different B cells compartments, the significant differences were primarily observed when comparing to synovial memory B cells and blood switched memory B cells from seropositive RA, whilst there was no significant difference compared to non-ACPA synovial plasma cells (Supporting Information Fig. 7A, B). Interestingly, also the HIV-derived sequences also displayed negative selection pressure in the CDRs that were significantly different from control antibodies in both VH and VL sequence analysis (Supporting Information Fig. 8).
When analyzing selection strengths solely based on glycosylation status, RA antibodies with N-glyc sites displayed significantly lower selection pressures in the heavy chain analysis compared to those without N-glyc motifs (Supporting Information Fig. 7C). Notably, the RA mAbs without glycosylation sites exhibit positive selection pressure in their VH CDRs, more in line with antigen-driven responses whereby the CDRs are to a larger extent responsible for the antigen-binding interactions (Supporting Information Fig. 7C).

### ACPA are more negatively charged and have a lower isoelectric point than control antibodies

Structure analysis comparing the ACPA clones with their predicted germline-rearranged sequences revealed that the numerous somatic hypermutations, which lie outside of the potential binding region, did not seem to dramatically change the over all structure. Instead, these contribute to promoted a change in surface charge towards more negatively charged acidic surface residues. ACPA had lower theoretical isoelectric point (PI) than their respectively germline-converted variable region sequences (Fig. 4B-D), observed in all clones bar two (37CEPT02C04 and 37CEPT01G09). The ACPA clones also had significantly lower PI than non-ACPA RA synovial mAbs as well as HIV bnAb and Pf memB clones (Fig. 4F). Isoelectric focusing (IEF) of CCP2-purified polyclonal ACPA revealed a trend for an overall shift towards lower PI compared to the CCP2 column flow-through and purified commercial polyclonal IgG (Fig. 4E). Even though polyclonal IgG appeared as a smear in IEF, as expected, this PI-shift could readily be detected when quantifying the color intensity by image analysis (Supporting Information Fig. 9). IEF of monoclonal antibodies correlated well with the theoretical PI of the antibodies, whereby the mAbs with lowest VH/VL PI were the ones that were separated at a lower pH on the gel and vice versa. The gels showed that the ACPA had a wide range of protein charge, generating PIs outside of the average pH 8.0-9.0 for human IgG1. Notably, the ACPA all had multiple isoform bands, consistent with post-translational modification and a high content of negatively charged sialic acid, while the control IgG displayed fewer bands (Fig. 4F, Supporting Information Fig. 9). We also had access to a genetically modified version of one ACPA (1325:01B09-mod). In this ACPA, framework hypermutations (including N-glyc sites in VH and VL) were converted back to the closest germline gene, whilst the CDRs were unchanged. Consequently, the 1325:01B09-mod did not contain N-glyc sites, but had an identical theoretical PI compared to the original 1325:01B09 clone, and our analysis demonstrated that it had maintained citrulline binding (Supporting Information Fig. 10). IEF showed that the modifications resulted in a higher actual PI and a reduction of isoform bands compared to the original clone (Fig. 4E).
From these results, we hypothesize that surface charge may be an important selection pressure in ACPA, and that negatively charged sialic acid-containing glycans could significantly contribute to such a non-classical BCR selection process. Together, this data suggests a potential for non-antigen driven mechanisms in the activation of ACPA+ B cells featuring BCR N-glyc motifs.

## Discussion

Our comprehensive single-cell sequence analysis revealed that Fab glycosylation is a prominent feature of ACPA+ B cells also when comparing to other chronic immune responses. N-glyc motifs in the variable region sequences of ACPA were acquired solely through somatic hypermutation, indicating that these are generated through germinal center responses. By comparing with antigen-specific Ig-sequences from HIV and *P. falciparum* malaria, we concluded that N-glyc levels were significantly increased in ACPA following SHM rate normalization. Consequently, anti-citrulline autoreactivity had a unique glycosylation profile compared to the control groups. These data are in line with and extend previous observations of Fab glycosylation in polyclonal anti-citrullinated protein responses in RA, non-expressed BCR sequences from CCP-sorted cells, and a limited number of expressed monoclonals [22, 26, 27]. In the current study, we also found that ACPA Fab glycans in the recombinant IgG were distinctly different from the Fc glycans and highly sialylated, consistent with what has been determined for Fab glycans of serum polyclonal antibodies in RA patients [23, 26]. Notably, a subset of the investigated mAbs have recently been found to have substantial in vitro and in vivo functionality and mediate enhanced osteoclastogenesis, pain behavior and endotoxin-induced arthritis [28, 29]. Therefore, the highlighted properties of Fab glycosylation and low charge of these 14 recombinant ACPA presented here within, may be important for future functional investigations.

The biological significance of hypersialylated ACPA in RA pathogenesis is still unknown. As decreased sialylation of the Fc-region has been found to increase the osteoclast-promoting properties of ACPA [30], it will be of interest to determine how Fab-associated sialylation may modulate the functional properties of anti-citrulline responses. Furthermore, the mechanism(s) driving the enzymatically mediated extent, and composition, of Fab glycosylation in Ig remains unclear. The finding that polyclonal ACPA purified from RA patient synovial fluid featured a higher content of Fab glycans compared to those derived from the blood [26], and the reported changes to the Fc-glycan composition of polyclonal ACPA in patients during disease progression [31], suggest that inflammation may regulate Ig glycosylation. Factors such as IL-6 and progesterone can modulate IgG glycosylation by affecting glycosyltransferases [32]. Furthermore, via an IL-23-Th17 cell axis, reduced expression of β-galactoside α2,6-sialyltransferase-1 in B cells has been observed in an inflammatory milieu [33].

Arguably, exactly what regulates the introduction of N-glyc motifs within BCR sequences and the selection pressures for N-glyc positive B cells remains even more elusive than the enzymatic pathways. Intriguingly, elevated Fab glycosylation has also been reported in vasculitis and primary Sjögren's Syndrome (pSS) [25, 34], which may imply that this phenomenon associates with autoimmunity, although this has yet to be determined. Our studies show that N-glyc rates were increased in synovial PC mAbs from seropositive patients even though the expressed mAbs showed no reactivity to citrullinated antigens. This indicates that there may be microenvironmental influences specific for seropositive RA contributing to driving the Fab glycosylation. However, we observed no significant difference in the presence of Fab N-glyc motifs within ACPA isolated from the synovium or blood.

The high levels of both N-glyc sites and SHM in ACPA suggest the potential for antigen-driven selection pressures in ACPA+ B cells. Yet, whether or not variable region glycosylation affects antigen binding remains unclear, as some studies report increases and decreases in binding [13, 35, 36], whilst others show no effect [37]. In our studies, we did not observe any effects on binding to citrullinated antigens. Structural modeling suggested that this may to some extent be explained by the observation that many of the N-glyc motifs are positioned outside of the predicted antigen-binding site. Therefore, a plausible explanation is that the acquisition of N-glyc in ACPA-Fab is mediated by non-(auto)antigen driven mechanisms. This is especially apparent when considering that the ACPA sequences showed low positive selection pressures in their CDRs, despite the extremely high levels of mutations. Furthermore, altered selection mechanisms have been previously suggested for autoimmune diseases such as RA and pSS [25, 38]. It is tempting to speculate that the selection of ACPA+ B cells could be mediated by the presence of Fab glycans in addition to the affinity for cognate antigen. Our structural analysis revealed that the extensive hypermutation outside of the CDRs of ACPA altered the charge of the variable regions towards more negatively charged surfaces, which was supported by a decreased variable region PI and overall charge of both polyclonal and monoclonal ACPA. Intriguingly, it has been reported that the addition of sialic acid glycans can also significantly decrease the PI of proteins [39] and that polyclonal IgG glycosylation in RA patients feature high acidic content [23]. Therefore, increases in Fab sialylation may augment a reduction of surface charge by SHM-driven mechanisms. As changes in surface charge may affect antigen interactions and modulate physical properties such as solubility and thermostability, it can be hypothesized that selection pressure for surface properties also outside of the binding site may be important in chronic responses. These have previously been neglected, and are not generally considered in selection pressure models based on replacement and silent mutations in CDRs compared to frame work regions. The predominance of negatively charged antibodies within the APCA responses may also potentially significantly influence the functional properties of soluble ACPA-IgG as well as result in increased circulation half-life and tissue retention due to reported altered interaction with the neonatal Fc-receptor [40].

The introduction of N-glyc motifs within variable regions has also been reported in non-Hodgkin's lymphoma [16, 41]. In the lymphoma studies, interesting alternative selection mechanisms have been extensively reported, including suggested lectin driven interactions with the observed high level of BCR Fab-glycan [14, 16, 41, 42]. It is well established that RA patients have an increased risk of developing lymphoma [43], which is linked to disease severity and high inflammatory activity [44]. Patients with pSS also have a higher risk of lymphoma [45] and, intriguingly, infiltrating B cells in parotid glands in SS feature a higher frequency of Fab glycosylation [25]. Therefore, it would be of interest to determine if autoimmune B cells and certain lymphoma B cells are driven by similar selection pressures, and if variable region glycosylation are related to the risk of B cells transforming into lymphoma cells.

A surprising finding was that HIV broadly-neutralizing antibodies (bnAb) also feature increased N-glyc motifs within their variable regions. To our knowledge, this is the first report of variable region N-glyc sites in antibodies derived from chronic infection, although the extent of Fab glycosylation in expressed HIV bnAbs remains to be determined. Despite chronic HIV infection and chronic autoimmunity featuring different pathogenesis, both involve constant antigen exposure and long-time immune activation. We speculate that antigen variation in both RA and HIV leads to the high mutation rates observed in the HIV bnAb, as well as the broadly cross-reactive ACPA.

In summary, our study highlights unique features of ACPA and explores the importance of variable region glycosylation in autoimmune disease and chronic inflammation. Given that glycosylation and negative surface charge are such strong characteristics of the anti-citrulline immune response, we hypothesize that acquisition of BCR N-linked glycosylation during the continuous evolution of autoreactive B cells may be important in disease pathogenesis, possibly mediating the breach of tolerance and/or expansion of autoreactive cells in the transition between latent autoimmunity and clinical active disease. Hence, these results may contribute to a better understanding of situations promoting Fab glycosylation, and its impact on the biological role of secreted ACPA and ACPA+ B cells.

## Materials and Methods

### *Patients samples*

All patients fulfilled the 2010 ACR/EULAR classification criteria for RA [2], and were classified as ACPA seropositive or seronegative disease using the IgG anti-CCP2 assay. All subjects gave informed consent and the study was approved by local ethics review committees. Synovial fluid samples were collected when the patients required arthrocentesis due to local disease activity. Bone marrow was obtained at the time of joint-replacement surgery. Mononuclear cells were prepared from samples with Ficoll-Paque separation (GE Healthcare) and cryopreserved until use.

### *Single B cell sorting and cloning*

Single B cells from different B cells populations and compartments were isolated as described below. All flow cytometry sorting was performed in accordance with the guidelines described in Cossarizza et al [46] and representative gating is shown in Supporting Information Fig. 11 and Fig. 12.

*RA bone marrow plasma cell derived clones:*

Single CD19+ (dim and bright) CD138+ plasma cells from bone marrow were sorted into 96-well PCR plates by flow cytometry.

*RA synovial memory B cells:*

Single CD19+ IgG+ memory B cells from synovial fluid were sorted by flow cytometry as previously described [47].

*RA synovial plasma cell derived ACPA positive and negative clones:*

For the isolation of single plasma cells from synovial fluid, the fluorescent foci method was applied [48], using anti-human IgG-specific beads and FITC-labelled anti-human IgG to identify IgG-secreting cells by fluorescent microscopy [28], which were extracted using an Eppendorf NK micromanipulator. Following cloning and expression, the ACPA positive clones were identified (Table 2).

*RA blood memory derived ACPA negative and positive clones obtained by tetramer sorting:*
Single citrulline-specific B cells were sorted by flow cytometry from peripheral blood of RA patients using an antigen-tetramer system based on phycoerythrin (PE)-conjugated streptavidin (Prozyme) and biotin cit-peptides [29]. Briefly, citrullinated human filaggrin (cfc1-cyc) or α-enolase (CEP-1) tetramer-specific cells were enriched using PE-MACS (Miltenyi Biotec) followed by flow cytometry sorting of CD19+ PE-tetramer+ B cells. Decoy-tetramers in PE-Alexa647 using arginine version of the peptides were utilized to remove unspecific cells. Recombinant mAbs were expressed and ACPA positive clones were identified (Table 2). Control CD19+IgD-CD27+ cells from the MACS flow through fraction was sorted as RA blood switched memory B cells.

*RA synovial memory derived ACPA positive clone:*
RA synovial CD27+ memory B cells were sorted with flow cytometry and transduced with Bcl-6 and Bcl-xL as previously described [49]. The transduced cells were sorted into wells with flow cytometry and cultured in the presence of CD40L and IL-21. Supernatants were screened for citrullinated peptide reactivity by ELISA and the ACPA clone 1003:10A01 was identified [50].

*RA blood memory derived ACPA positive clone obtained by EBV immortalization:*
The ACPA positive blood memory derived clone BCVA1 was isolated from RA PBMC using methods previously described [51, 52]. Briefly, CD22+ cells were enriched using MACS microbeads (Miltenyi) and subsequently sorted for IgG+ cells. The cells were immortalized with EBV, seeded into 384 wells, and cultured in the presence of 2.5 μg/ml CpG (ODN 2006) and irradiated allogenic mononuclear cells. Supernatants were screened for reactivity to citrullinated vimentin and cDNA was generated.

Cloning and expression of the synovial memory B cell-derived clones have been previously reported [47, 53, 54]. Immunoglobulin variable genes were cloned from single B cells into human heavy- and light-chain IgG expression vectors, after cDNA synthesis and PCR amplification, using established methods [47, 55]. All clones were subsequently expressed at the Karolinska institute using a standardized protocol and screened for cit-peptide reactivity by ELISA and multiplex microarray [56].

***Antibody production and purification***
Monoclonal antibodies were produced as hIgG1 by transient co-transfection of heavy- and light-chain vectors by PEI-max (Polysciences Inc.) into Expi293 cells (Life Technologies) and affinity purified by Protein G Fast-Flow Sepharose (GE Healthcare). Antibody concentrations were determined by anti-human IgG ELISA. IgG proteins were assessed by SDS-PAGE, SEC analysis for aggregation, and antigen-specificity ELISA.

***Ig gene sequence analysis***
N-glyc motifs were identified in translated sequences of V-(D)-J regions using the NetNGlyc1.0 server (www.cbs.dtu.dk/services/NetNGlyc/), which screens sequences for the consensus N-linked glycosylation motif (N-*X*-S/T, whereby *X* is any amino acid except proline). Ig-sequences were analyzed by alignment to germline sequences using IMGT/V-quest to confirm if N-glyc were introduced through SHM [17]. The BASELINe program (Bayesian estimation of Ag-driven selection), which evaluates SHM patterns in Ig sequences to generate quantitative visualization of selection pressures [24]. Two-point table formats reveal selection pressure comparisons between groups, using both statistically derived numerical values and probability distribution graphs.

***Detection of N-glycans in monoclonal IgG.***
Glycans were cleaved from 12 ACPA mAbs using PNGase F (NEB) or Endo S (IgGZERO, Genovis), followed by separation of 1 μg IgG on SDS-PAGE, 4-12% Bis-Tris gel (Life Technologies), and staining with SimplyBlue™ SafeStain (Life Technologies). For lectin immunoblots, IgG were transferred to a PVDF membrane (Life Technologies), blocked with Carbo-Free blocking solution (Vector Laboratories) and glycans were detected with biotin-SNA or biotin-ConA (Vector Laboratories) followed by anti-biotin-HRP (Jackson ImmunoResearch) and development with Clarity™ Western ECL substrate (BioRad). For lectin-ELISAs, 96-well ½ area high binding plates (Corning) were coated with goat $(Fab')_2$ anti-human IgG (Fab) capture antibody (Jackson ImmunoResearch), blocked with Carbo-Free blocking solution, and incubated with mAbs in Carbo-Free blocking buffer. Reactivity was detected with biotin-SNA (2 μg/ml) or biotin-ConA (0.2 μg/ml) and streptavidin-HRP (Jackson ImmunoResearch), and developed with TMB substrate (Biolegend). Isoelectric focusing (IEF) of polyclonal and monoclonal ACPA was achieved using Novex pH 3-10 gels and Novex IEF buffer system (Life Technologies) following the manufacturer's instructions with separation of 10 μg polyclonal IgG or 5 μg monoclonal IgG at 100V 1hr, 200V 2hrs, and 500V 40min. Gels were fixed in 12% trichloroacetic acid and stained with SimplyBlue™ SafeStain.

***Antigen-binding assay***

The ability of monoclonal ACPA to bind citrullinated peptides was assessed by Quanta Lite CCP3 IgG ELISA at indicated concentrations according to the manufacturer's instructions (Inova Diagnostics).

***Glycopeptide profiling via liquid chromatography - mass spectrometry analysis (LC-MS/MS)***

Mass spectrometry was performed on native or desilaylated or degalactosylated ACPA mAbs for an in-depth analysis of Fab glycosylation. IgG samples (15 μg/sample) were trypsin digested as previously described [8]. Samples were kept at 10°C and injected onto the chromatographic column in 1 μg aliquots.

Glycosidase treatment was performed over night at 37°C on trypsin digests using 2.5 mU of 1) α(2–3,6,8,9)-Sialidase A and 2) α(2–3,6,8,9)-Sialidase A and β(1–3,4)-Galactosidase (ProZyme, Hayward, CA) dissolved in 10 μL of the corresponding X10 buffers provided by the manufacturer.

A nano-liquid chromatography system Ultimate 3000 connected in-line to an Elite Orbitrap mass spectrometer (ThermoFisher Scientific) was used. Reversed phase LC-separation of the peptides was performed on a 50 cm long EASY spray column (PepMap, C18, 3 μm, 100 Å) using a gradient solvent system containing (A) water with 2% acetonitrile and 0.1% formic acid and (B) acetonitrile with 2% water and 0.1% formic acid. The gradient was set up as follows: 1–30% (B) in 79 min, 31–95% (B) in 5 min, 95% (B) for 8 min and 1% (B) for 10 min. The flow rate was set at 300 nL/min. The mass spectrometer was operating in the positive DDA mode with a survey mass spectrum in the range of m/z 150-2000 with a nominal resolution of 60,000. Following each MS scan, top eight most abundant precursor ions were selected for MS/MS with collision induced dissociation (CID) and electron transfer dissociation (ETD).

***Glycopeptide identification and quantitation***

Fc-IgG-glycopeptide amino acid sequences and glycoforms were characterized as previously described [8, 57]. Briefly, $IgG_1$ Fc-glycopeptides were identified by their characteristic retention times (as determined by IgG standard) and accurate monoisotopic masses (within <10 ppm from the theoretical values) of doubly and triply charged ions (EEQYNSTYR) as well as of triply and quadruply charged ions (TKPREEQYNSTYR).

The IgG Fab glycans of monoclonal IgG were identified by the predicted tryptic peptides with N-linked glycosylation sites in the heavy and light chain sequences. The specific Fab glycans were identified via accurate monoisotopic masses (within <10 ppm from the theoretical values), MS/MS fragmentation and sialidase and galactosidase treated samples (Supporting Information Fig. 7).

Quantification of Fc and Fab glycopeptides and the corresponding unmodified peptide sequences were performed in a label-free manner using Quanti similarly to what has previously been described [8, 58]. Glycopeptide ion abundances were integrated over respective chromatographic monoisotopic ion peaks (<10 ppm from the theoretical values) at the charged states described above and within a ±1.5 min interval around the expected retention times. Abundances of the Fc-glycopeptides and the light and heavy chain glycopeptides were normalized by their respective total content. A list of glycan compositions with corresponding suggested structures and relative distributions on respective glycopeptide is given in Supporting Information Table 3.

***Molecular modelling***

VH-VL structure models were generated using the Immunoglobulin Structure (PIGS) web server and the best H and L chain method [59], and visualized by Jmol (Jmol: an open source Java viewer for chemical structures in 3D, www.jmol.org/). The GlyProt server was utilized for *in silico* glycosylation using a basic glycan structure [60]. Molecular electrostatic potential (MEP) calculations were calculated with the PDB2PQR server [61] and Jmol MEP surface using RWB color scheme (scale -0.5, 0.5). Theoretical isoelectric points (PI) for VH-VL were calculated with the ExPaSY protparam online tool.

***Statistical analysis***

Statistical analysis was performed in Prism (GraphPad) using Kruskal-Wallis test with Dunn's correction for multiple comparison or 2-sided Mann-Whitney test for comparing continuous variables between groups, as appropriate. Frequencies were compared using Fisher's exact test. A $p<0.05$ was considered significant.

## Acknowledgements

We would like to thank Dr Fredrik Wermeling (Karolinska Institutet) for reagents and valuable discussions, Dr Yan Wang (Karolinska Institutet) for sequence analysis advice, and Ragnhild Stålesen and Dr Monika Hansson (Karolinska Institutet) for support in antibody production, validation and characterization. We thank Dr Christian Busse (German Cancer Research Center) for kindly providing the paired V-region sequences from B cells isolated from malaria patients. We also thank UCB Pharma for support with technology and antibody production as well as financial support for Katy Lloyd's postdoctoral training within the IMI BTCure project. This work was supported by the Swedish Research Council, the Swedish Rheumatism Association, King Gustaf V's 80-year Foundation, and the IMI EU funded project BeTheCure 115142.

## Conflict of interest

The authors declare no commercial or financial conflict of interest.

**Table 1. Details of N-linked glycosylation in monoclonal ACPA variable region sequences.**

| ACPA | Predicted N-glyc sites | B cell type | V region N-glyc | N-glyc aa motif | Germline aa motif† | Location | Confirmed glycosylation in recombinant IgG# |
|---|---|---|---|---|---|---|---|
| 1325:01B09 | 1VH, 1VL | Synovial plasma cell | VH | NGS | TGT | FWR1 | SDS-PAGE, lectin ELISA, MS** |
| | | | VL | NTS | SSS | FWR1 | SDS-PAGE, MS** |
| 1325:04C03 | 2VH | Synovial plasma cell | VH | NMT | TMT | FWR3 | SDS-PAGE, lectin ELISA, SNA blot** |
| | | | | NFS | SFS | CDR3 | |
| 1325:05C06 | 2VH | Synovial plasma cell | VH | NVS | TVS | FWR1 | SDS-PAGE, lectin ELISA, SNA blot |
| | | | | NLT | KLS | FWR3 | |
| 1325:07E07 | 2VH, 1VL | Synovial plasma cell | VH | NDT | SSS | CDR1 | SDS-PAGE, lectin ELISA, MS |
| | | | | NSS | NPS | FWR3 | |
| | | | VL | NLT | TLT | FWR3 | SDS-PAGE, SNA-blot, MS |
| 37CEPT02C04 | 3VH, 1VL | Blood IgG+ memory, CEP1 sorted | VH | NVS | TVS | FWR1 | SDS-PAGE, lectin ELISA, SNA-blot |
| | | | | NGS | GGS | CDR1 | |
| | | | | NVS | NVY | CDR3 | |
| | | | VL | NIS | TIS | FWR1 | SDS-PAGE, SNA-blot |
| 37CEPT01G09 | 2VH | Blood IgG+ memory, CEP1 sorted | VH | NAS | SCY | CDR3 | SDS-PAGE, lectin ELISA, SNA-blot |
| | | | VH | NVS | TVS | FWR1 | |
| 14CFCT03F09 | 1VL | Blood IgG+ memory, CFC1 sorted | VL | NFT | TIT | FWR1 | N.D |
| 14CFCT02E04 | 1VH | Blood IgG+ memory, CFC1 sorted | VH | NVS | TVS | FWR1 | N.D |
| 14CFCT02D09 | 1VL | Blood IgG+ memory, CFC1 sorted | VL | NRS | NKN | CDR1 | lectin ELISA€, SNA-blot |
| 14CFCT02H12 | 1VL | Blood IgG+ memory, CFC1 sorted | VL | NNS | KNY | CDR1 | lectin ELISA€, SNA-blot |
| 14CFCT03G09 | 2VH | Blood IgG+ memory, CFC1 sorted | VH | NVS | DTS | FWR3 | SDS-PAGE, lectin ELISA, SNA-blot** |
| | | | | NLT | KLS | FWR3 | |
| 62CFCT01E04 | 2VH, 1VL | Blood IgG+ memory, CFC1 sorted | VH | NTT | SSS | CDR1 | SDS-PAGE, lectin ELISA, SNA-blot |
| | | | | NVT | SVT | FWR3 | |
| | | | VL | NFT | DFT | FWR3 | SDS-PAGE, SNA-blot |
| BVCA1 | 1VL | Blood IgG+ memory | VL | NGS | SGS | FWR3 | Double bands, lectin ELISA, SNA blot |
| 1003:10A01 | 1VL | Synovial IgG+ memory | VL | NNS | SSS | CDR3 | SDS-PAGE, lectin ELISA, SNA-blot |
| | | | | NST | SST | CDR3 | |

† As determined by IMGT V-QUEST [17]. *N.D.* not determined, due to low recombinant expression yields
# Recombinant human IgG1, determined by digestion with endoglycosidases (PNGase and EndoS) and separation by SDS-PAGE, and SNA blots, and/or ConA/SNA lectin ELSA and mass spectrometry. Notably the ELISA cannot determine if the glycosylation is localized on the heavy or light chain
** Large batch variations were observed for sialic acid content

**Table 2. N-linked variables region glycosylation sites in ACPA compared to antibodies derived from other B-cell groups.**

| | VH SHM* | VH N-glyc sites/SHM | | VL SHM* | VL N-glyc sites/SHM | | Total paired VH+VL N-glyc sites/SHM | | Total frequency N-glyc positive clones | | |
|---|---|---|---|---|---|---|---|---|---|---|---|
| | *Mean* | *Mean±SD* | *p-value#* | *Mean* | *Mean±SD* | *p-value#* | *Mean±SD* | *p-value#* | *% (n/N)* | *p-value€* | *OR (95% CI)* |
| **ACPA** (N=14) | 52 | 0.027±0.025 | | 34 | 0.027±0.029 | | 0.032±0.018 | | 100% (14/14) | | |
| ***Control mAb sequences*** | | | | | | | | | | | |
| Healthy memB (N=27) | 22 | 0.007±0.018 | **0.0002** | 14 | 0.002±0.008 | **<0.0001** | 0.004±0.0097 | **<0.0001** | 19% (5/27) | **<0.0001** | 118 (6.1-2313) |
| HIV bnAb (N=19) | 82 | 0.005±0.009 | **0.02** | 70 | 0.007±0.010 | **0.03** | 0.006±0.008 | **0.001** | 63% (12/19) | **0.01** | 17.4 (0.9-336) |
| Pf memB (N=103) | 29 | 0.009±0.020 | **<0.0001** | 19 | 0.005±0.016 | **<0.0001** | 0.007±0.013 | **<0.0001** | 25% (26/103) | **<0.0001** | 84.8 (4.9-1472) |
| ***RA mAb sequences*** | | | | | | | | | | | |
| memB, CCP+ RA (N=25) | 24 | 0.071±0.016 | **0.0009** | 20 | 0.0095±0.20 | **<0.0001** | 0.008±0.013 | **<0.0001** | 32% (8/25) | **0.001** | 59.7 (3.2-1126) |
| BM PC, CCP+ RA (N=198) | 30 | 0.006±0.017 | **<0.0001** | 22 | 0.003±0.014 | **<0.0001** | 0.005±0.010 | **<0.0001** | 21% (42/198) | **<0.0001** | 107 (6.2-1828) |
| ***RA non-ACPA expressed syn. mAbs§*** | | | | | | | | | | | |
| syn. PC, CCP- RA (N=27) | 22 | 0.004±0.015 | **<0.0001** | 16 | 0 | **<0.0001** | 0.002±0.007 | **<0.0001** | 7% (2/27) | **<0.0001** | 79.0 (4.2-1501) |
| syn. PC, CCP+ RA (N=33) | 21 | 0.016±0.023 | **0.05** | 15 | 0.003±0.012 | **<0.0001** | 0.009±0.012 | **<0.0001** | 39% (13/33) | **<0.0001** | 44 (2.4-802) |
| syn. memB, CCP+ RA (N=20) | 24 | 0.006±0.016 | **0.0005** | 9 | 0 | **<0.0001** | 0.003±0.008 | **<0.0001** | 15% (3/20) | **<0.0001** | 145 (6.9-3045) |

Healthy memB: Blood memory B cell derived mAbs from a healthy individual, HIV bAb: HIV-derived broadly neutralizing antibodies; Pf memB: *Plasmodium falciparum specific* mAbs; mem, CCP+ RA: blood derived memory B cells from CCP+ RA; BM PC, CCP+ RA: bone marrow plasma cell derived mAbs from CCP+ RA; syn. PC, CCP- RA: synovium plasma cell derived expressed mAbs from RA; syn. PC, CCP- RA: synovium plasma cell derived expressed mAbs from CCP + RA; syn memB, CCP+ RA. synovium memory B cell derived expressed mAbs from CCP+ RA.

Predicted N-linked glycosylation sites were identified in the variable regions of immunoglobulin genes from single-cell cloned B cells using the NetNGlyc1.0 server (www.cbs.dtu.dk/services/NetNGlyc/), searching for classical N-X-S/T sites.

n= number of clones with variable region glycosylation site

N= total number of clones in the B-cell group

§ All non-ACPA synovial mAbs were expressed and extensively evaluated and showed no reactivity to citrullinated antigens.

* All antibodies in the B cell groups included in the analysis were selected to have more than 15 SHM in the heavy chain and/or the light chain for more accurate comparison to ACPA sequences.

# P-value compared to the ACPA group derived from Kruskal-Wallis test with adjustment for multiple comparisons.

€ P-value compared to the ACPA group derived from Fisher's exact test

**Table 3. Mass spectrometry analysis determined the glycan composition of two ACPA mAbs.**

| Monoclonal IgG | **1325:01B09*** | | | **1325:07E07**[§] | | |
|---|---|---|---|---|---|---|
| IgG region | Fc (YN**S**T**Y) | HC (L**N**C**S**V) | LC (G**N**T**S**N) | Fc (Y**N**S**T**Y) | HC (I**N**D**T**T) | LC (A**N**L**T**I) |
| Total number of identified glycopeptides (n) | 21 | 39 | 48 | 16 | 33 | 40 |
| Number of identified sialylated glycopeptides (n) | 2 | 18 | 25 | 1 | 5 | 19 |
| Relative abundance of sialylated glycoforms /peptide (%) | 0.1% | 49% | 60% | 0.1% | 3% | 63% |

A full list of the identified glycan compositions and relative abundances is given in Supporting Information Table 4.
* 1325:01B09 batch C was used for analysis, previously confirmed to feature high content of sialylation (Supporting Information Fig. 4)
§ 1325:07E07 batch B was used for analysis

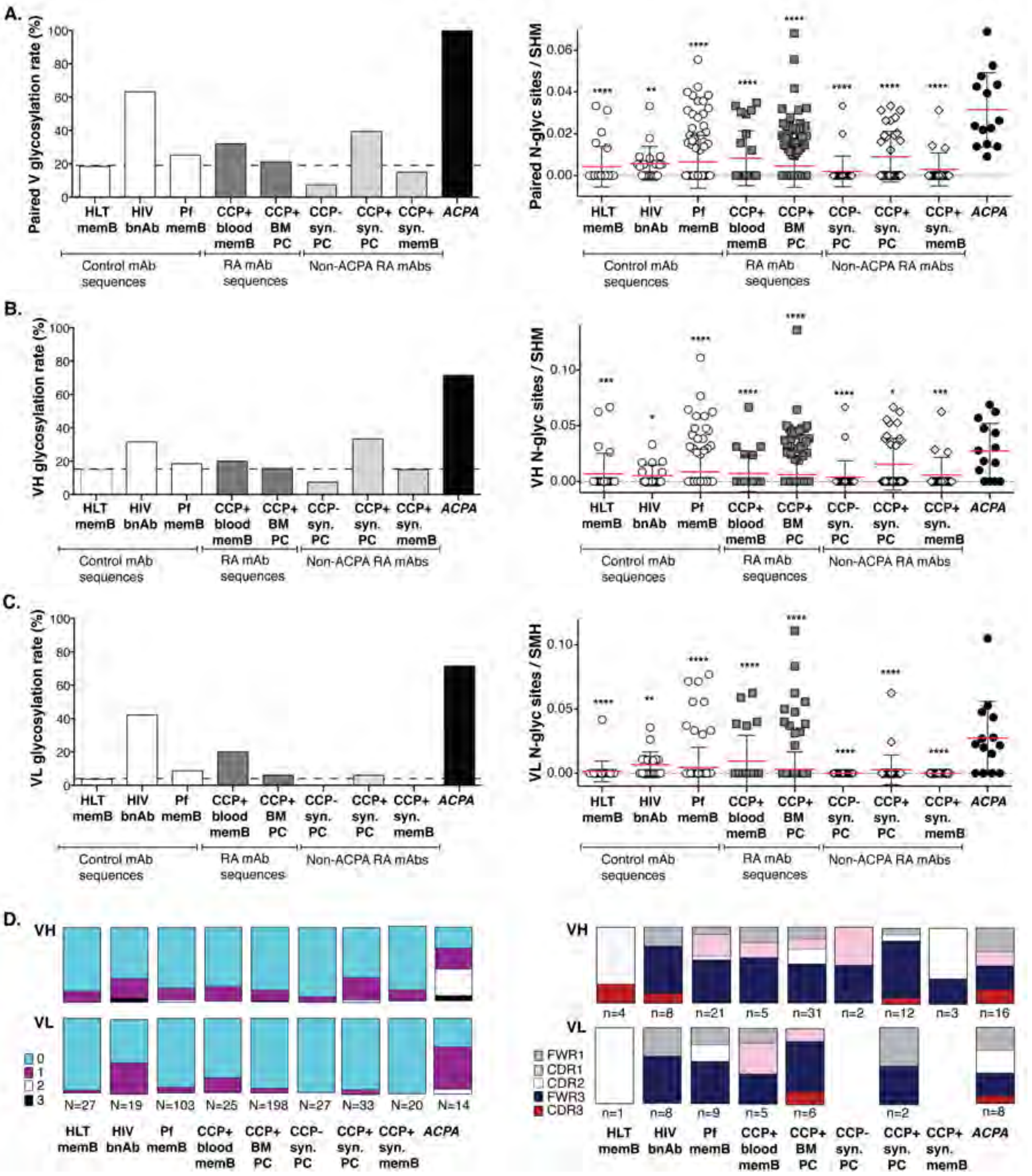

**<u>Figure 1</u> Monoclonal ACPA-IgG contain significantly increased N-glyc motifs.**
Consensus N-glyc motifs (N-*X*-S/T) were identified in paired V chain (**A**) VH chain (**B**), and VL chain (**C**) sequences. The analysis compares RA ACPA mAbs with sequences from 27 heathy circulating memory B cells (healthy memB), 27 synovial plasma cells from seronegative RA (CCP- syn. PC), and 25 circulating memory B cells (CCP+ RA blood memB), 198 bone marrow plasma cells (CCP+ RA BM PC), 33 synovial plasma cells (CCP+ RA syn. PC), and 20 synovial memory B cells (CCP+ RA syn. memB) from seropositive RA patients. The analysis also included published broadly-neutralizing HIV antibodies (HIV bnAb) [18] and *Plasmodium falciparum* antibodies (Pf memB) [19]. All clones were selected to carry more than 15 somatic hypermutations (SHM) in either the VH or VL sequences. Glycosylation rates were depicted as the percentage of sequences containing N-glyc motifs (left panels) or N-glyc rates normalized for SHM (right panels). P-values were determined by Kruskal-Wallis test with adjustment for multiple comparisons, * p-value <0.05, ** p-value<0.01, *** p-value<0.001 **** p-value<0.0001. Mean and SD are depicted. **D**. Number of N-glyc sites per V-chain (left panel) and location of N-glyc sites within VH and VL chains (right panel) in the framework regions (FWRs) and complementary-determining regions (CDRs). BCR variable region sequences were analyzed *in silico* using IMGT V-quest [17] and the NetNGlyc1.0 server (www.cbs.dtu.dk/services/NetNGlyc/).

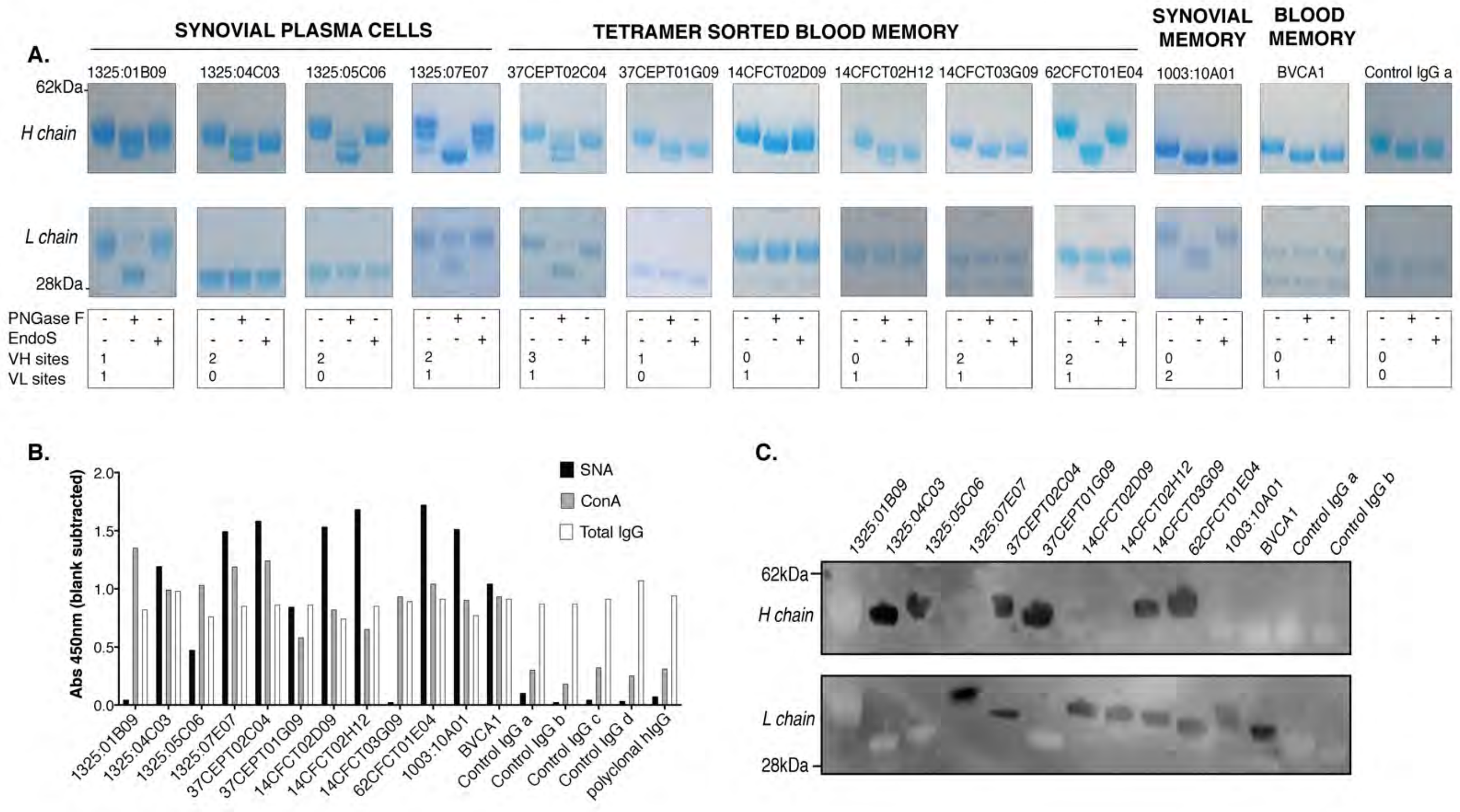

**Figure 2** **Deglycosylation revealed that Fab glycan motifs are occupied in monoclonal ACPA.**

**A**. Coomassie blue stained SDS-PAGE of 12 recombinantly expressed monoclonal ACPA-IgG treated with PNGase F (cleaves all N-linked glycans) and Endo S (cleaves only Fc N-linked glycans) or left untreated, run under reducing conditions. Glycan compositions were determined by lectin ELISA (**B**) and by SNA immunoblotting detecting sialic acid containing glycans (**C**), which revealed high rates of sialylation and mannosylation for the ACPA but not the control mAb from RA synovial memory B cells (Control IgG(a) 1362:01E02, Control IgG(b) 1276:01G09), control IgG from RA plasma cell (Control IgG(c) 1276:06D06) or the commercial human IgG1 isotype control (Control IgG(d) ET901, Biolegend). Data shown in A, B and C are representative of three independent experiments. ELISA reactivity in B was analyzed in duplicates.

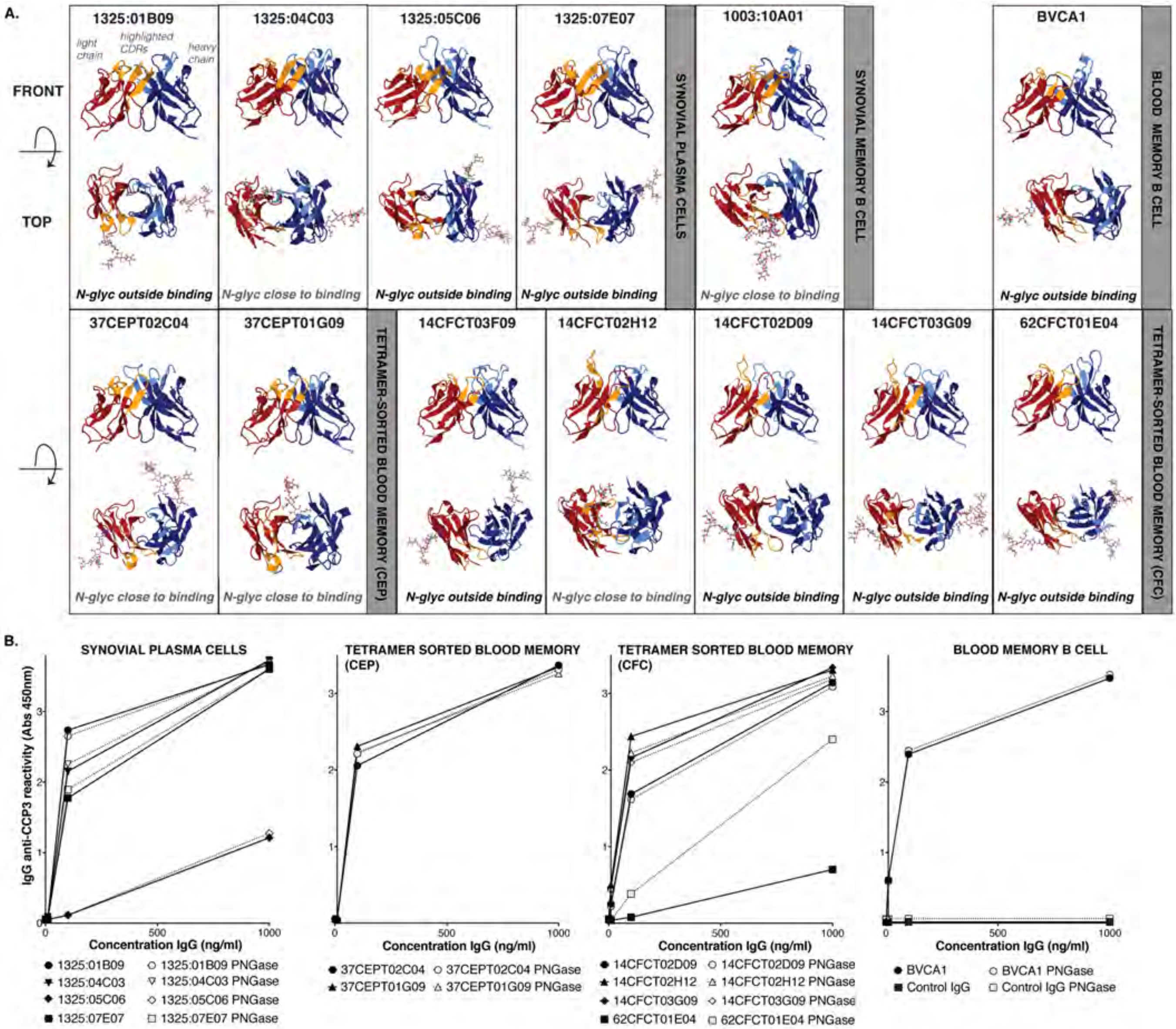

**<u>Figure 3</u> Fab glycosylation did not affect antigen binding.**

**A.** Homology-based structural predictions of ACPAs with predicted N-glyc sites revealed that Fab N-glyc sites on ACPA are primarily positioned outside of the predicted antigen-binding surface built up by the CDR, whereas others lie close to potential binding regions. Models are visualized for the VH/VL region of the mAbs as cartoons of light chain variable region in red and heavy chain region in blue with the light chain CDR loops highlighted in orange and heavy chain CDR loops in light blue. The VH/VL region are shown in a "top" view, looking down into the potential binding site generated by the CDR loops. Structure models were generated with PIGS online tool using best H + L model, basic glycan structures were added with GlyProt server and visualized with Jmol. **B**. The removal of glycans did not significantly alter ACPA binding to CCP3 by ELISA (Quanta Lite CCP3 IgG modified protocol, Inova Diagnostics). The mAb 1003:10A01 is not positive for CCP3 binding and therefore excluded in this assay. The recombinantly expressed synovial memory B cell derived clone 1362:01E02 was used as isotype control.

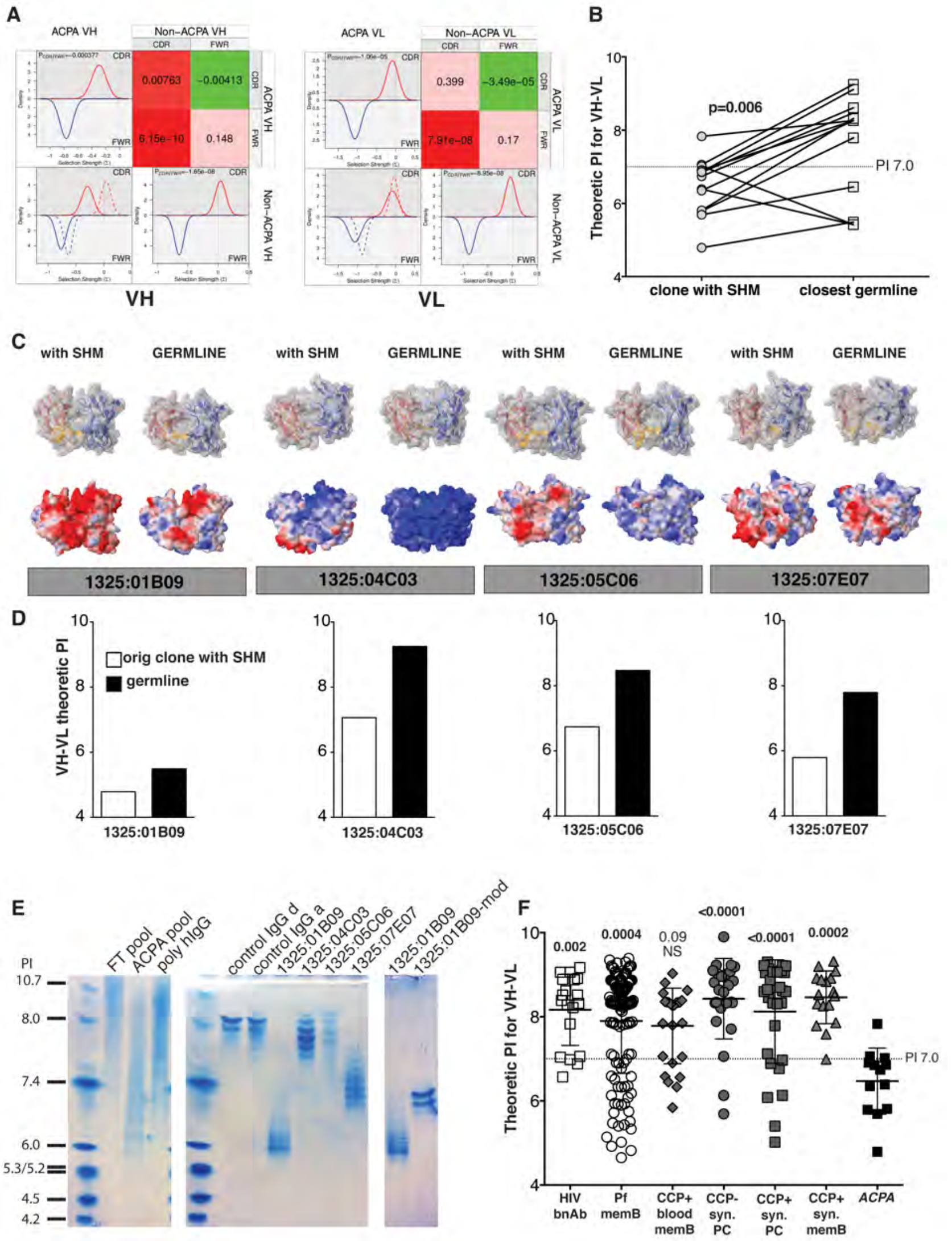

**<u>Figure 4</u> ACPA show distinct differences in selection pressure and surface charge compared to other antibody groups**

**A.** The predicted selection pressures were obtained for ACPA sequences compared to non-ACPA RA (synovial memory, synovial plasma cells and blood memory B cells from seropositive RA) using BASELINe computational tool [24], whereby selection pressures for the CDR and the FWR (upper left and lower right plots, respectively) and overlays (lower left plot) are visualized for heavy chains (left panel) and light chains (right panel). The dotted line depicts the row group and the solid line represents the column group. Positive selection pressures are depicted as red color and positive values, whereas negative selection pressures are shown as green color and negative values. The intensities of the colors depict the strength of the relative selection pressures. **B**. Theoretical isoelectric points (PI) for VH-VL of 13 ACPA and their predicted germline rearrangement. P-value from pair t-test is presented. **C**. Homology-based structural surface predictions of four plasma cells derived ACPA and their predicted germline rearrangements showing electrostatic potential (MEP) surfaces. The RWB color (scale -.5, .5) shows positively charged surfaces as blue and negatively charged as red. The theoretical PI for the clones are depicted in (**D**). **E.** Isoelectric focusing (IEF) of human RA-derived IgG. The left panel shows a shift towards lower PI in IEF of polyclonal CCP2 affinity purified ACPA IgG (ACPA pool) compared to CCP2 column flow through (FT pool) and polyclonal IgG. The right panels show IEF of recombinant human monoclonal IgG1, revealing heterogeneity in PI of ACPA, low PI and multiple isoforms. Genetic modification of the ACPA clone 1325:01B09 to remove the two N-glyc sites in the VH and VL framework resulted in an increase in PI and a reduction of isoforms. The non-ACPA plasma cell derived clone 1276:06D06 and the memory B cells derived clone 1362:01E02 were used as controls in IEF. **F**. Theoretical VH-VL PI for different antibody groups compared to ACPA. P-values were determined by Kruskal-Wallis test with adjustment for multiple comparisons, mean and SD are shown for each group.

**Supporting Information Table 1. Cloned and screened clones from single B cells from RA patients and controls.**

| Patient ID | Recruiting hospital | Status | Sex | Age | Source of B cells | B cell subset | No of clones* |
|---|---|---|---|---|---|---|---|
| RA146 | University Hospitals Birmingham | CCP+RA | F | 69 | Synovial tissue | non-ACPA syn. memory | 4 |
| RA153 | University Hospitals Birmingham | CCP+RA | F | 40 | Synovial fluid | non-ACPA syn. memory | 2 |
| RABVCA | Geneva University Hospital | CCP+RA | N/A | N/A | Blood | ACPA | 1 |
| RA1003 | Karolinska University Hospital | CCP+RA | F | 67 | Synovial fluid | non-ACPA syn. memory | 2 |
| RA1276 | Karolinska University Hospital | CCP+RA | F | 68 | Synovial fluid | ACPA<br>non-ACPA syn. memory<br>non-ACPA syn. PC | 1<br>6<br>6 |
| RA1325 | Karolinska University Hospital | CCP+RA | F | 58 | Synovial fluid | non-ACPA syn. memory<br>non-ACPA syn. PC<br>ACPA | 2<br>25<br>4 |
| RA1444 | Karolinska University Hospital | CCP+RA | M | 51 | Synovial fluid | non-ACPA syn. memory<br>non-ACPA syn. PC | 4<br>2 |
| RA268 | Karolinska University Hospital | CCP+RA | F | 83 | Bone marrow | BM PC | 198 |
| RA1362 | Karolinska University Hospital | CCP-RA | M | 66 | Synovial fluid | CCP neg syn. PC | 27 |
| RA37 | University of Minnesota Medical Center | CCP+RA | F | 51 | Blood | blood memory | 9 |
| RA14 | University of Minnesota Medical Center | CCP+RA | F | 37 | Blood | ACPA (CEP-sorted)<br>blood memory<br>ACPA (CFC-sorted) | 2<br>2<br>3 |
| RA62 | University of Minnesota Medical Center | CCP+RA | F | 67 | Blood | blood memory<br>ACPA (CFC-sorted) | 14<br>1 |
| HC14 | Karolinska University Hospital | Control | F | 54 | Blood | healthy memory | 27 |

* All analyzed clones were selected to have >15 SHM in the heavy chain and/or light chain

**Supporting Information Table 2. HIV broadly neutralizing monoclonal antibodies containing N-linked glycosylation motifs within variable region sequences.**

| bnAb | Variable region | N-glyc motif # | Germline motif † | Domain † | References |
|---|---|---|---|---|---|
| PG9 | VL | NGT | TGT | FWR1 | [1] |
| VRC01 | VL | NLT | TFT | FWR3 | [2] |
| VRC02 | VL | NLT | TFT | FWR3 | [2] |
| PGT121 | VH | NYS | NYN | FWR3 | [3] |
| | | NLS | TIS | FWR3 | |
| | | NGT | NGT | FWR3 | |
| PGT122 | VH | NVS | TVS | FWR1 | [3] |
| PGT125 | VH | NVS | TVS | FWR1 | [3] |
| | VL | NGT | TGT | FWR1 | |
| PGT130 | VH | NVT | SVT | FWR3 | [3] |
| | VL | NGT | TGT | FWR1 | |
| PGT131 | VH | NVT | SVT | FWR3 | [3] |
| 8ANC195 | VH | NLT | AED | FWR3 | [4] |
| 3BNC117 | VL | NLT | TFT | FWR3 | [4] |
| 3BNC60 | VL | NLT | TFT | FWR3 | [4] |
| NIH45-46 | VL | NLS | TLT | FWR3 | [4] |

# As determined using the NetNGlyc1.0 server (http://www.cbs.dtu.dk/services/NetNGlyc/), to predict N-linked glycosylation motifs (N-X-S/T) within the variable region amino acid sequences of translated Ig genes.

†As determined by IMGT V-QUEST [5].

**Supporting Information Table 3. Relative abundance and the glycan compositions identified of two ACPA mAbs**

| **Glycan** | | **1325:01B09*** | | | **1325:07E07§** | | |
|---|---|---|---|---|---|---|---|
| **Suggested structure#** | **Composition** | **Fc** (Y**N**S**T**Y) | **HC** (L**N**C**S**V) | **LC** (G**N**T**S**N) | **Fc** (Y**N**S**T**Y) | **HC** (I**N**D**T**T) | **LC** (A**N**L**T**I) |
| GlcNAc | HexNAc[1] | 0.2% | 0.1% | 1% | 0.5% | - | 0.1% |
| M3 | HexNAc[2]Hex[3] | - | 0.1% | - | - | - | 0.2% |
| M4 | HexNAc[2]Hex[4] | 0.3% | - | - | 0.1% | - | - |
| M5 | HexNAc[2]Hex[5] | 1% | 1% | 2% | 0.3% | 7% | 0.6% |
| M6 | HexNAc[2]Hex[6] | - | - | - | - | - | - |
| M7 | HexNAc[2]Hex[7] | - | - | - | 0.1% | 0.2% | - |
| A1 | HexNAc[3]Hex[3] | 1% | 0.9% | - | 1% | 0.1% | 2% |
| A2 | HexNAc[4]Hex[3] | 0.2% | 7% | - | 0.2% | - | 14% |
| A2B or A3 | HexNAc[5]Hex[3] | 0.5% | 2% | 0.2% | 2% | 0.3% | 3% |
| A4 | HexNAc[6]Hex[3] | - | - | 0.2% | - | 0.5% | 0.3% |
| FGlcNAc | dHex[1]HexNAc[1] | - | - | 0.3% | 0.1% | - | - |
| FM3 | dHex[1]HexNAc[2]Hex[3] | - | - | 0.2% | - | 0.6% | - |
| FM6 | dHex[1]HexNAc[2]Hex[6] | - | - | - | - | 0.1% | - |
| FM7 | dHex[1]HexNAc[2]Hex[7] | - | - | 0.2% | - | - | - |
| FM9 | dHex[1]HexNAc[2]Hex[9] | - | - | 0.2% | - | 0.8% | - |
| FM10 | dHex[1]HexNAc[2]Hex[10] | - | - | 0.1% | - | 0.1% | - |
| FA1 | dHex[1]HexNAc[3]Hex[3] | 7% | 0.4% | 1% | 6% | 6% | - |
| FA2 | dHex[1]HexNAc[4]Hex[3] | 65% | 6% | 9% | 73% | 17% | 0.2% |
| FA2B or FA3 | dHex[1]HexNAc[5]Hex[3] | 0.2% | 3% | 9% | 0.2% | 22% | 0.2% |
| FA4 | dHex[1]HexNAc[6]Hex[3] | - | 0.1% | 1% | - | 17% | - |
| A1G1 | HexNAc[3]Hex[4] | 0.1% | 0.1% | - | - | - | 1% |
| M5A1 or | HexNAc[3]Hex[5] | 0.1% | - | - | - | - | - |
| M6A1 or | HexNAc[3]Hex[6] | 0.1% | - | - | - | 1% | - |
| A2G1 | HexNAc[4]Hex[4] | - | 5% | - | - | 0.5% | 6% |
| A2G2 | HexNAc[4]Hex[5] | - | 3% | - | - | 0.1% | 2% |
| A2BG1 or A3G1 | HexNAc[5]Hex[4] | 0.1% | 4% | - | 0.3% | - | 4% |

| Glycan | | 1325:01B09[*] | | | 1325:07E07[§] | | |
|---|---|---|---|---|---|---|---|
| **Suggested structure[#]** | **Composition** | **Fc** (Y**N**S**T**Y) | **HC** (L**N**C**S**V) | **LC** (G**N**T**S**N) | **Fc** (Y**N**S**T**Y) | **HC** (I**N**D**T**T) | **LC** (A**N**L**T**I) |
| A2BG2 or A3G2 | HexNAc[5]Hex[5] | - | 0.2% | - | - | - | 0.4% |
| A3G3 | HexNAc[5]Hex[6] | - | - | - | - | 0.5% | 0.1% |
| A4G1 | HexNAc[6]Hex[4] | - | 0.1% | 0.1% | - | 0.2% | 0.1% |
| A4G2 | HexNAc[6]Hex[5] | - | - | - | - | 0.2% | 0.1% |
| A4G3 | HexNAc[6]Hex[6] | - | - | 0.1% | - | 0.1% | - |
| FA1G1 | dHex[1]HexNAc[3]Hex[4] | 1% | 0.1% | - | 0.3% | - | - |
| FM5A1 or | dHex[1]HexNAc[3]Hex[5] | 0.6% | - | - | - | - | - |
| FM6A1 or | dHex[1]HexNAc[3]Hex[6] | 0.2% | - | - | - | 2% | - |
| FM7A1 or | dHex[1]HexNAc[3]Hex[7] | - | - | - | - | 1% | - |
| FM8A1 or | dHex[1]HexNAc[3]Hex[8] | - | - | - | - | 0.1% | - |
| FA2G1 | dHex[1]HexNAc[4]Hex[4] | 21% | 6% | 2% | 15.6% | 4% | 1% |
| FA2G2 | dHex[1]HexNAc[4]Hex[5] | 1% | 5% | 0.3% | 0.6% | 3% | 0.5% |
| FA2BG1 or | dHex[1]HexNAc[5]Hex[4] | 0.1% | 8% | 6% | - | 12% | 1% |
| FA2BG2 or | dHex[1]HexNAc[5]Hex[5] | - | 0.3% | 2% | - | 0.5% | 0.1% |
| FA3G3 | dHex[1]HexNAc[5]Hex[6] | - | - | 3% | - | 0.0% | - |
| FA4G1 | dHex[1]HexNAc[6]Hex[4] | - | - | 1% | - | 0.7% | - |
| FA4G2 | dHex[1]HexNAc[6]Hex[5] | - | - | 1% | - | - | - |
| FA4G3 | dHex[1]HexNAc[6]Hex[6] | - | - | 0.3% | - | - | - |
| M4A1G1S1 | HexNAc[3]Hex[5]NeuAc[1] | - | - | 0.1% | - | - | - |
| A1G1S1 | HexNAc[3]Hex[4]NeuAc[1] | - | - | - | - | - | 0.2% |
| A2G1S1 | HexNAc[4]Hex[4]NeuAc[1] | - | 2% | 0.4% | - | 0.2% | 7% |
| A2G2S1 | HexNAc[4]Hex[5]NeuAc[1] | - | 10% | 0.3% | - | - | 22% |
| A2G2S2 | HexNAc[4]Hex[5]NeuAc[2] | - | 4% | - | - | - | 21% |
| A2BG1S1 or | HexNAc[5]Hex[4]NeuAc[1] | - | 3% | - | - | - | 5% |
| A2BG2S1 or | HexNAc[5]Hex[5]NeuAc[1] | - | 0.2% | - | - | - | 1% |
| A2BG2S2 | HexNAc[5]Hex[5]NeuAc[2] | - | 0.2% | 0.4% | - | - | 0.2% |
| A3G3S1 | HexNAc[5]Hex[6]NeuAc[1] | - | 0.1% | - | - | - | 1% |
| A3G3S2 | HexNAc[5]Hex[6]NeuAc[2] | - | 0.2% | - | - | - | 1% |

| **Glycan** | | **1325:01B09*** | | | **1325:07E07§** | | |
|---|---|---|---|---|---|---|---|
| **Suggested structure#** | **Composition** | **Fc** (YNSTY) | **HC** (LNCSV) | **LC** (GNTSN) | **Fc** (YNSTY) | **HC** (INDTT) | **LC** (ANLTI) |
| A4G1S1 | HexNAc[6]Hex[4]NeuAc[1] | - | - | 0.2% | - | 0.1% | 0.1% |
| A4G2S1 | HexNAc[6]Hex[5]NeuAc[1] | - | 0.1% | 0.1% | - | - | 0.1% |
| A4G2S2 | HexNAc[6]Hex[5]NeuAc[2] | - | - | 0.1% | - | - | 0.1% |
| A4G3S2 | HexNAc[6]Hex[6]NeuAc[2] | - | - | 1% | - | - | - |
| A4G3S3 | HexNAc[6]Hex[6]NeuAc[3] | - | - | 0.2% | - | - | - |
| FM4A1G1S1 | dHex[1]HexNAc[3]Hex[5]NeuAc[1] | 0.1% | - | - | - | - | - |
| FA2G1S1 | dHex[1]HexNAc[4]Hex[4]NeuAc[1] | 0.1% | 2% | 2% | 0.1% | 1.5% | 0.1% |
| FA2G2S1 | dHex[1]HexNAc[4]Hex[5]NeuAc[1] | - | 14% | 7% | - | 1.5% | 3% |
| FA2G2S2 | dHex[1]HexNAc[4]Hex[5]NeuAc[2] | - | 8% | 9% | - | 0.2% | 0.3% |
| FA2BG1S1 or | dHex[1]HexNAc[5]Hex[4]NeuAc[1] | - | 4% | 6% | - | - | 1% |
| FA2BG2S1 or | dHex[1]HexNAc[5]Hex[5]NeuAc[1] | - | 0.3% | 4% | - | - | 0.1% |
| FA2BG2S2 | dHex[1]HexNAc[5]Hex[5]NeuAc[2] | - | - | 3% | - | - | - |
| FA3G3S1 | dHex[1]HexNAc[5]Hex[6]NeuAc[1] | - | 0.1% | 7% | - | - | 0.1% |
| FA3G3S2 | dHex[1]HexNAc[5]Hex[6]NeuAc[2] | - | 0.2% | 13% | - | - | 0.1% |
| FA4G1S1 | dHex[1]HexNAc[6]Hex[4]NeuAc[1] | - | - | 0.2% | - | - | - |
| FA4G2S1 | dHex[1]HexNAc[6]Hex[5]NeuAc[1] | - | 0.1% | 1% | - | - | - |
| FA4G2S2 | dHex[1]HexNAc[6]Hex[5]NeuAc[2] | - | - | 1% | - | - | - |
| FA4G3S1 | dHex[1]HexNAc[6]Hex[6]NeuAc[1] | - | 0.1% | 0.2% | - | - | - |
| FA4G3S3 | dHex[1]HexNAc[6]Hex[6]NeuAc[3] | - | - | 1.1% | - | - | - |
| FA4G4S3 | dHex[1]HexNAc[6]Hex[7]NeuAc[3] | - | - | 1.2% | - | - | - |
| FA4G4S1 | dHex[1]HexNAc[6]Hex[7]NeuAc[1] | - | - | 0.3% | - | - | - |
| FA4G4S2 | dHex[1]HexNAc[6]Hex[7]NeuAc[2] | - | - | 0.5% | - | - | - |

* 1325:01B09 batch C was used for analysis, previously confirmed to feature high content of sialylation (FigureS4)
§ 1325:07E07 batch B was used for analysis
# Nomenclature is according to

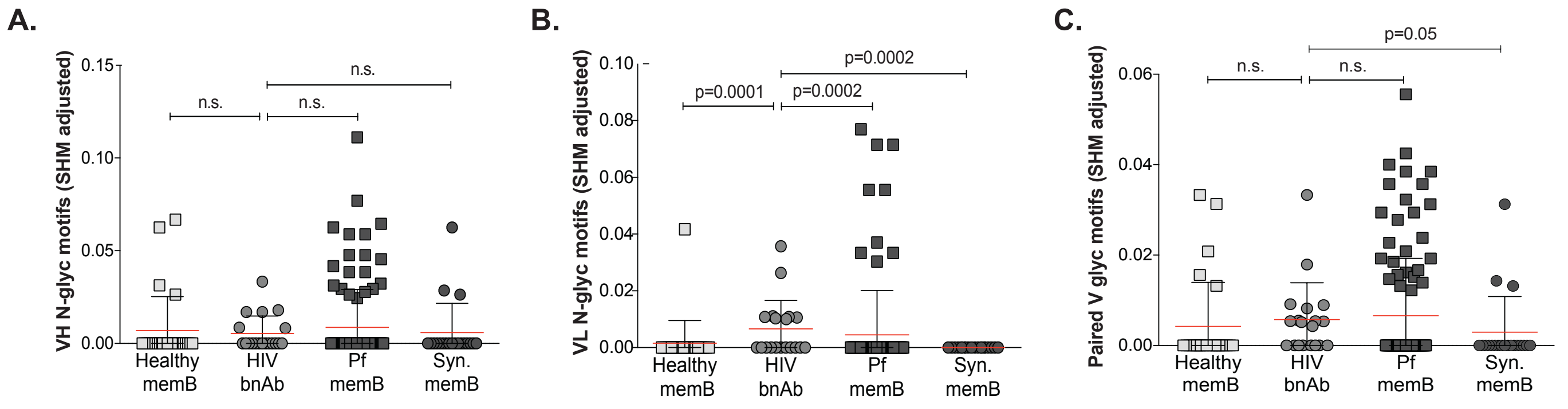

**Supporting Information Figure 1. N-glyc motifs are significantly increased in HIV bnAbs.**

**A.** VH chain memory B cell sequences revealed no significant differences in N-glyc rate, whereas rates were significantly increased in HIV bnAb VL chains **(B)** and paired V chains (**C**) compared to the other sequences. Statistical significance regarded as $p<0.05$, as determined by Kruskal-Wallis test.

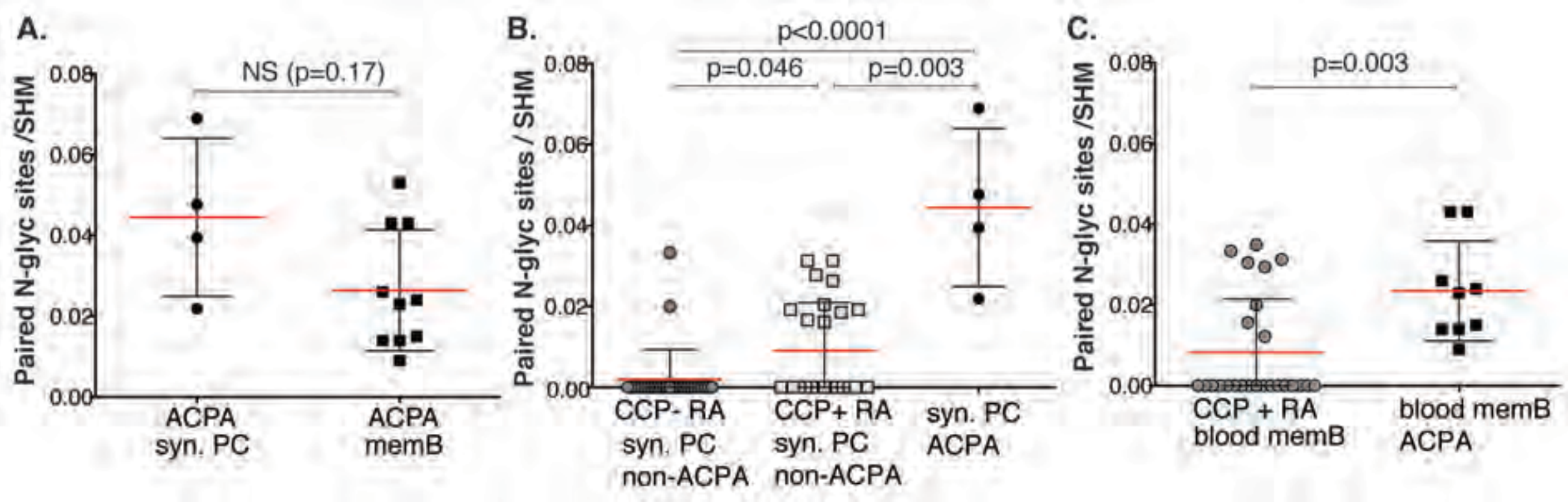

**Supporting Information Figure 2. N-glyc motif rates of mAbs isolated from different anatomical locations.**

**A.** Comparison of sequences showed no significant difference in N-glyc rate in ACPAs derived from synovial plasma cell (PC) or blood memory B cell (memB) **B.** N-glyc rates in expressed mAbs derived from synovial plasma cell with no ACPA reactivity (from CCP+ RA or CCP- RA) compared to identified ACPAs from plasma cells. **C.** N-glyc rate comparison between randomly selected blood switched memory B cells in CCP+ RA and identified ACPAs isolated from circulating memory B cells. All sequences were selected to have more than 15 SHMs in either the heavy chain or light chain. P-values were determined by Kruskal-Wallis or Mann-Whitney test as appropriate.

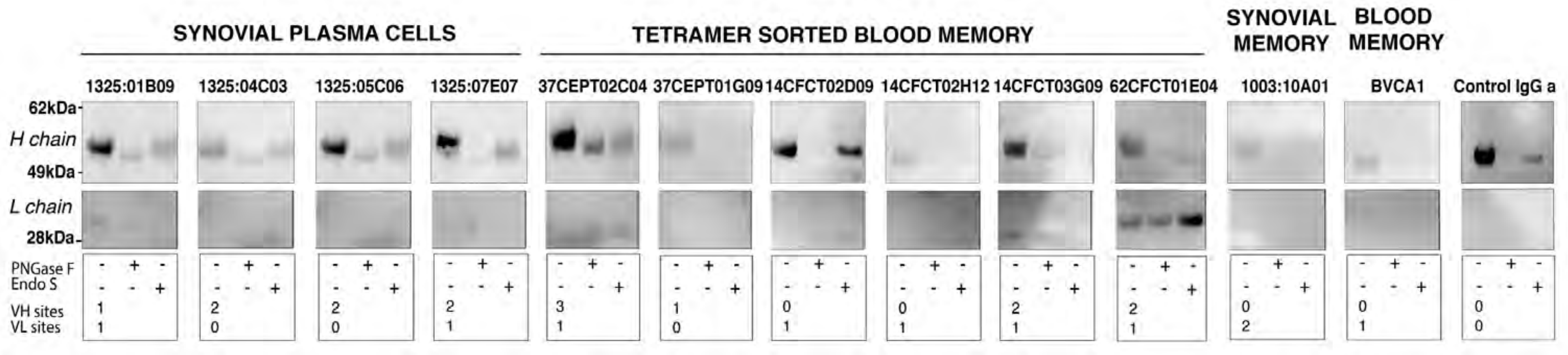

**Supporting Information Figure 3. Lectin blot confirmed deglycosylation of ACPA-IgG.**

ConA blotting of untreated, PNGase F-digested, and EndoS-digested mAbs (1μg hIgG1) revealed reduced binding of biotinylated ConA to endoglycosidase-cleaved ACPA. The predicted N-glyc motif numbers in the mAbs are also shown.

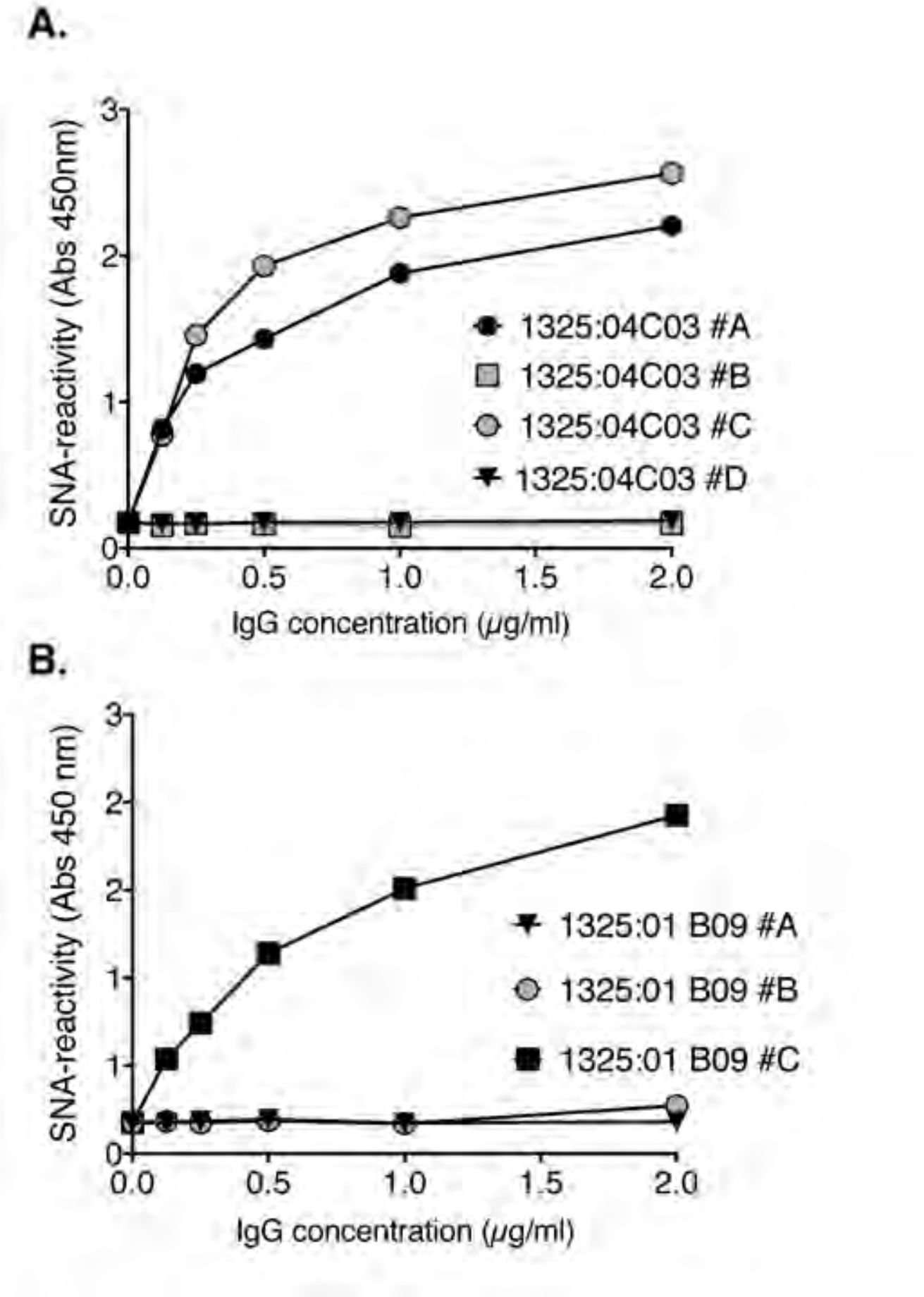

**Supporting Information Figure 4. Monoclonal ACPA exhibit batch variation in glycosylation rates.**

Lectin-ELISA showed that 1325:04C03 (**A**) and 1325:01B09 (**B**) batches had different levels of sialylation, as determined as reactivity against biotinylated SNA. Means of duplicates are shown.

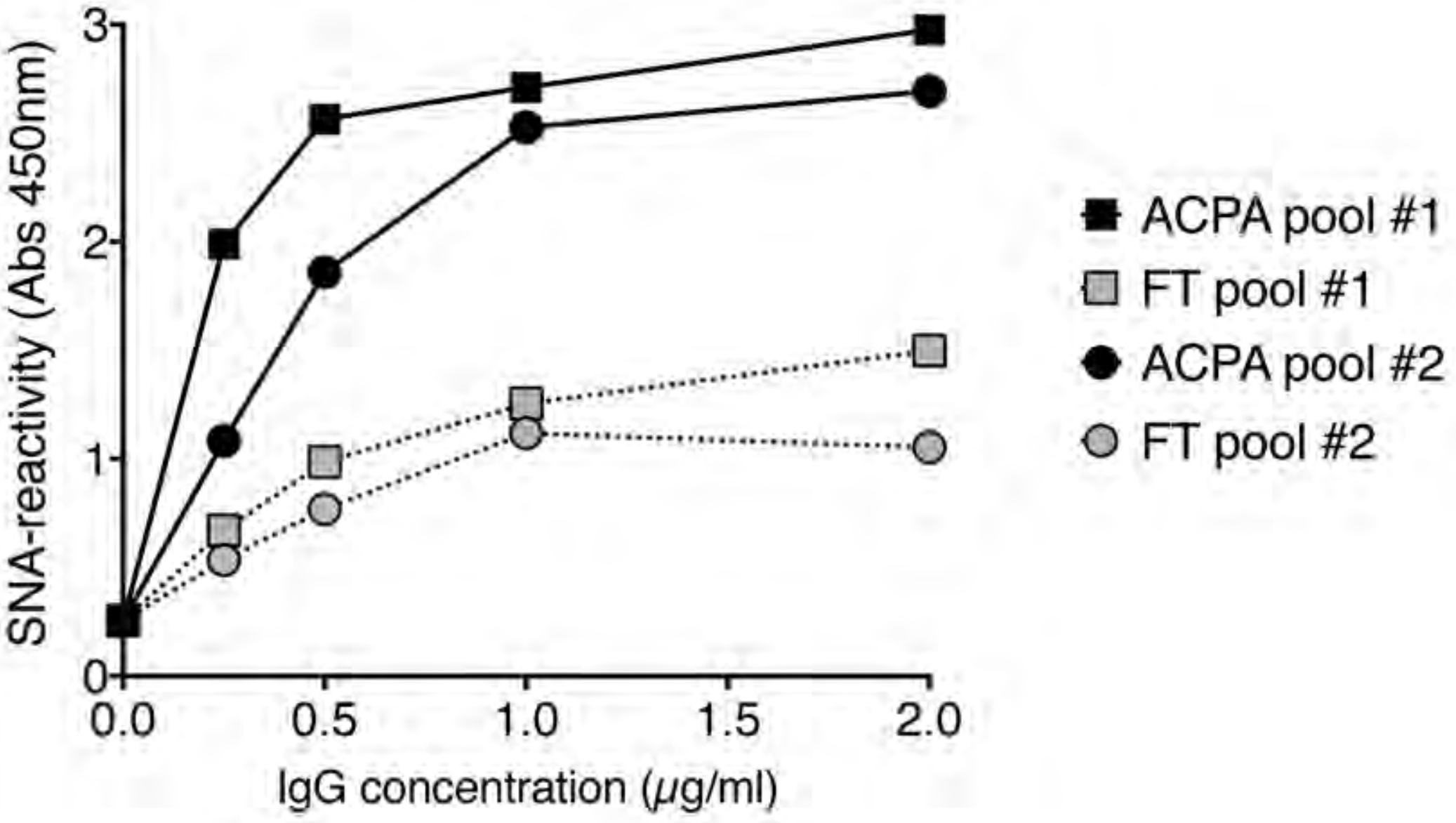

**Supporting Information Figure 5. Sialylation extent of polyclonal human ACPA**

Polyclonal serum affinity purified anti-CCP2 IgG (ACPA pool) contain a high level of sialylation compared to the flow through IgG from the CCP2 column (FT pool), measured by SNA lectin ELISA. Two independent ACPA pools were analyzed (#1 purified from 35 RA patients; #2 purified from 108 RA patients. Means of duplicates are show.

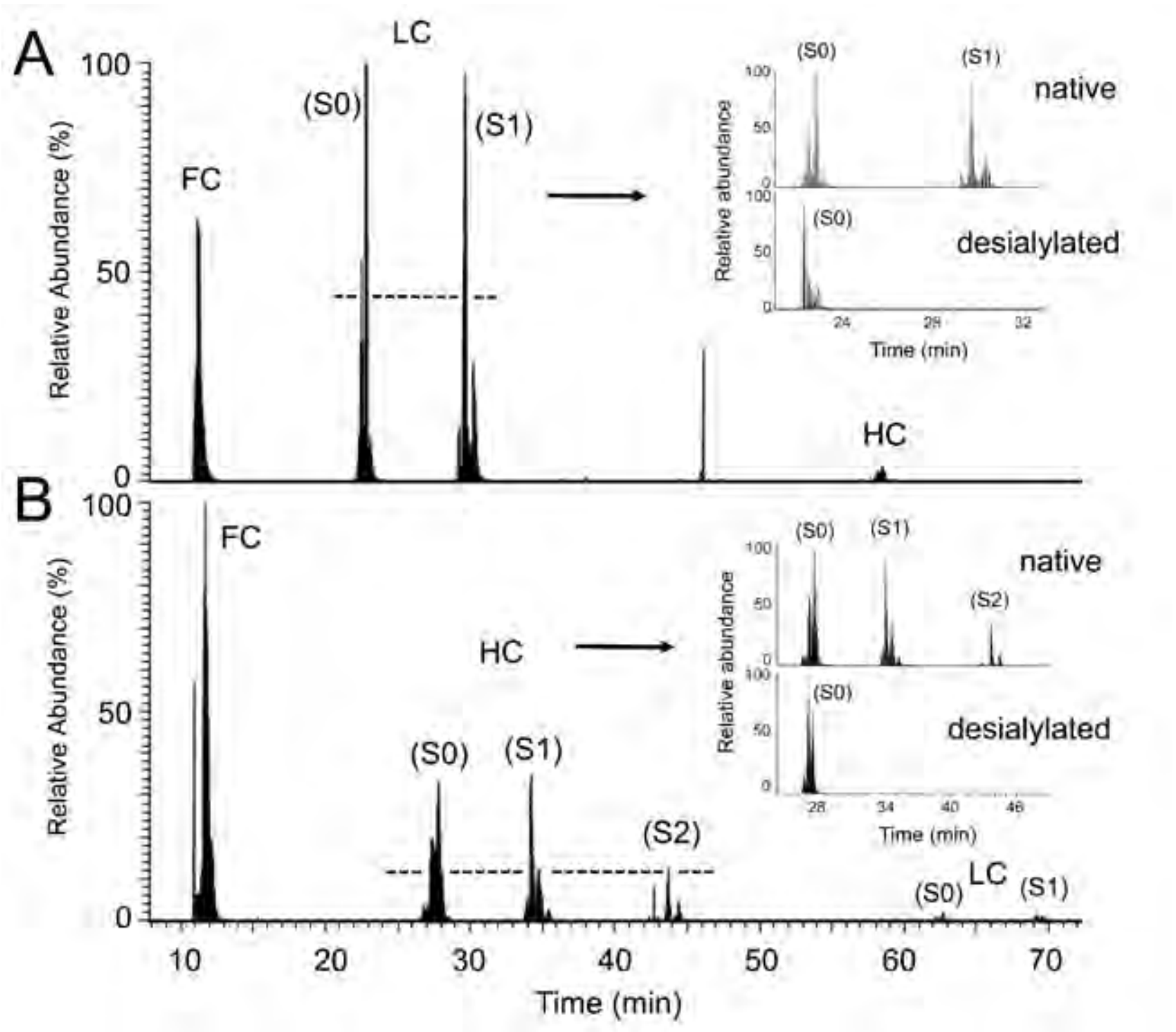

**Supporting Information Figure 6. Mass spectrometry analysis of ACPA glycopeptides**

Merged ion chromatograms of the monoisotopic glycopeptide ions. **A.** Merged ion chromatograms of glycopeptides from 1325:07E07. Main glycopeptides from the FC region (EEQYNSTYR), HC region (HLHLQESGPGLVKPSETLSLTCTVSGGSINDTTYYWGWIR) and LC region (SEDTANLTITR) are shown. Note that the glycoforms that does not contain sialic acid (S0) and the forms containing one sialic acid (S1) separate according to retention time. Following desialylation, the S0 forms remain intact but the desialylated forms are gone from the chromatogram. **B.** Merged ion chromatograms of glycopeptides from 1325:01B09. Main glycopeptides from the FC region (EEQYNSTYR), LC region (VTIPCSGNTSNIGYNIVNWYQQVPGTAPK) and the HC region (TSETLSLNCSVSR) are shown. Similarly, as for 1325:07E07, the glycoforms that does not contain sialic acid (S0) and the forms containing one sialic acid (S1) or two sialic acids (S2) separate according to retention time. Following desialylation, the S0 forms remain intact but the desialylated forms are gone from the chromatogram.

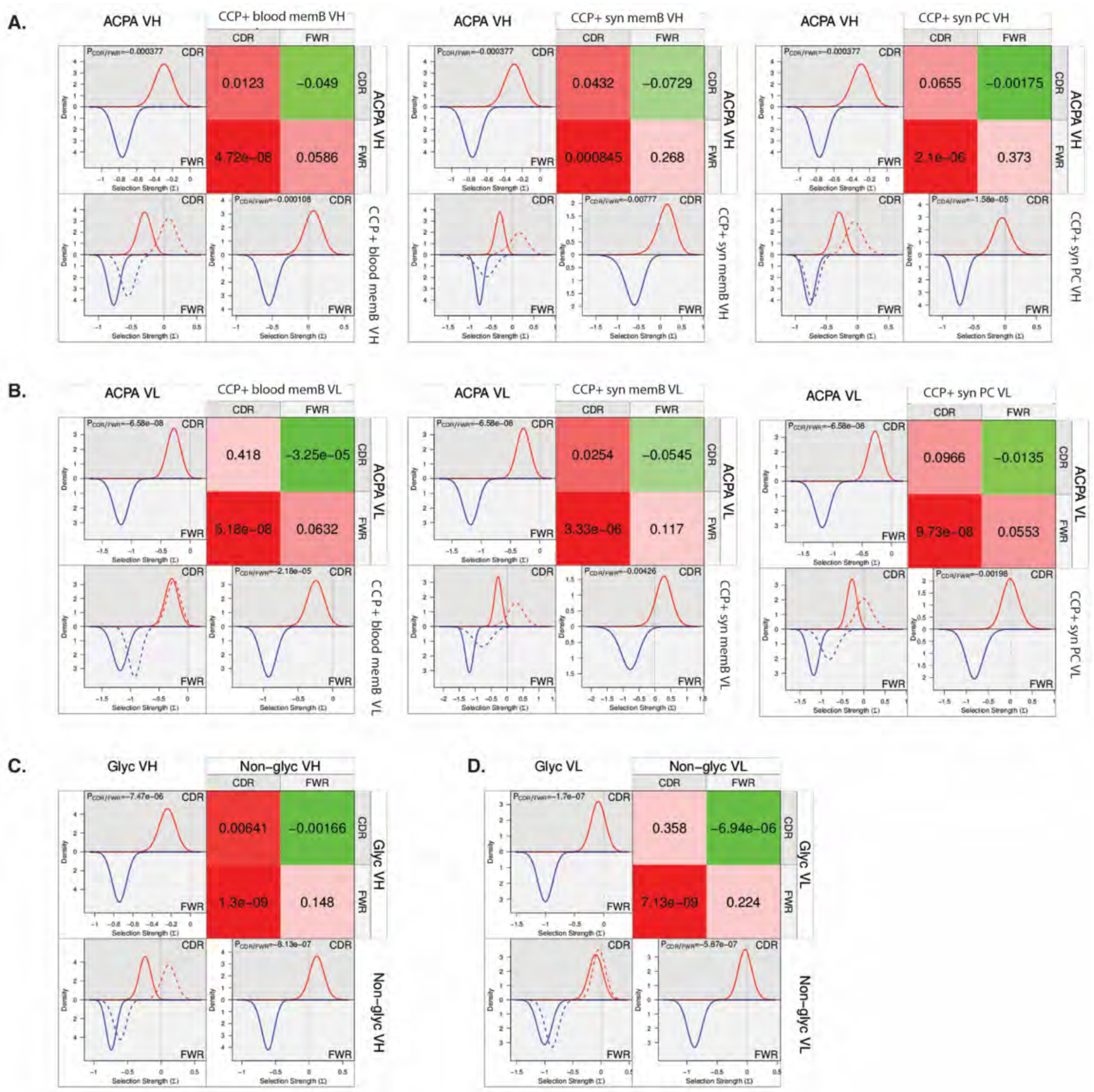

**Supporting Information Figure 7. Selection pressure in ACPA compared to non-ACPA RA derived mAbs**

Predicted selection pressures were obtained for ACPA versus different sub-groups of highly mutated non-ACPA, for the VH **(A)** and VL (**B**) sequences. Differences in the selection pressures in VH (**C**) or VL (**D**) sequences of RA-derived mAbs with or without N-glyc motifs were also compared. The figure shows output data from the BASELINe web tool [6]. Positive selection pressures are depicted as red color and positive values, whereas negative selection pressures are shown as green color and negative values. The graphs compare the distribution graphs of both groups, whereby the dotted line depicts the row group and the solid line represents the column group. Statistical differences were derived from binomial statistical tests by BASELINe, whereby the intensities of the colors depict the strength of the relative selection pressures.

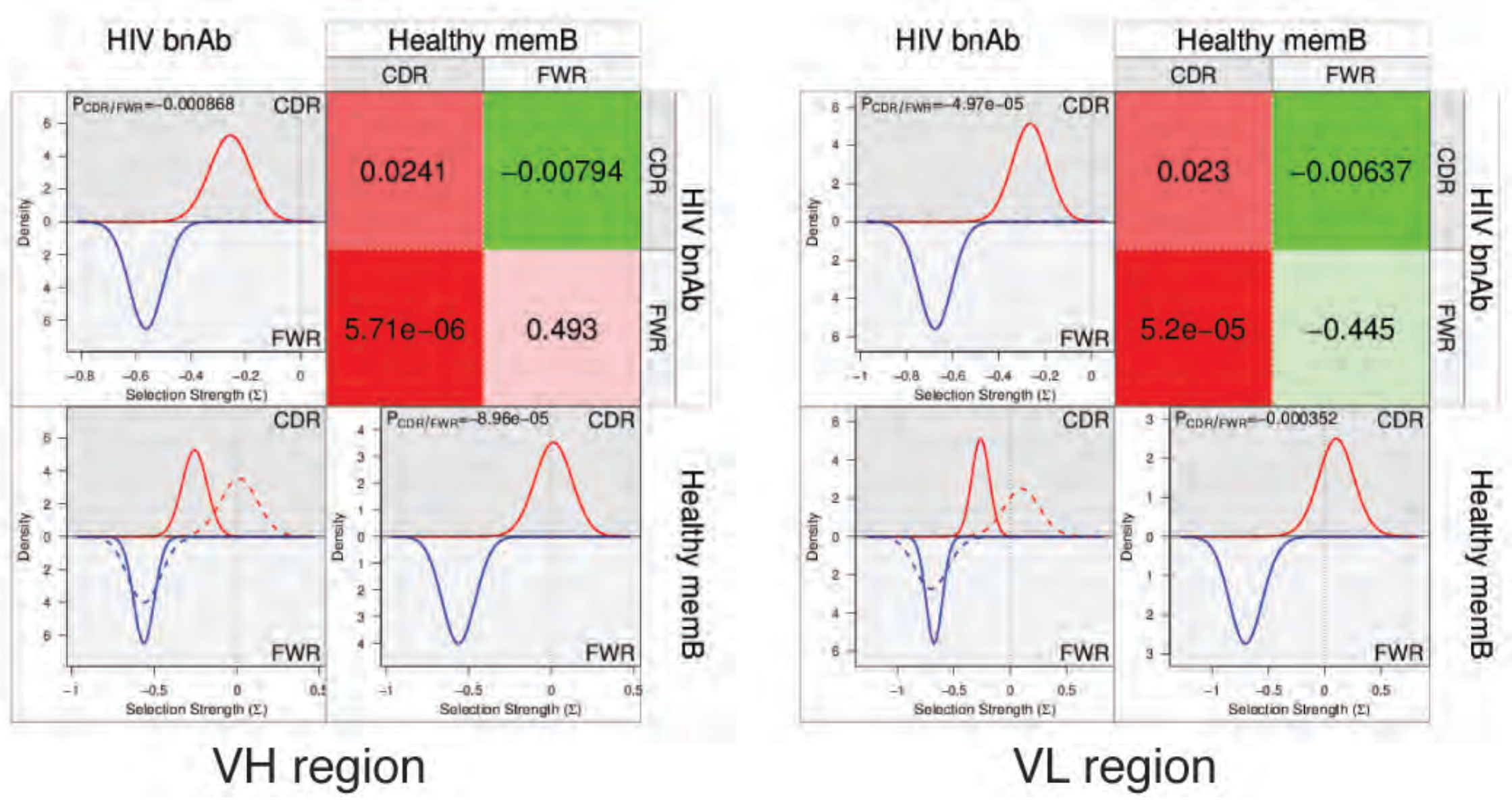

**Supporting Information Figure 8. Selection pressure in HIV broadly neutralizing mAbs compared to controls**

Predicted selection pressures were obtained for HIV broadly neutralizing antibodies (HIV bnAb) and circulating memory B cells from a healthy individual (Healthy memB). Selection pressures in the VH (left) and VL (right) regions are illustrated. The figure shows output data from the BASELINe web tool [6]. Positive selection pressures are depicted as red color and positive values, whereas negative selection pressures are shown as green color and negative values. Statistical differences were derived from binomial statistical tests by BASELINe.

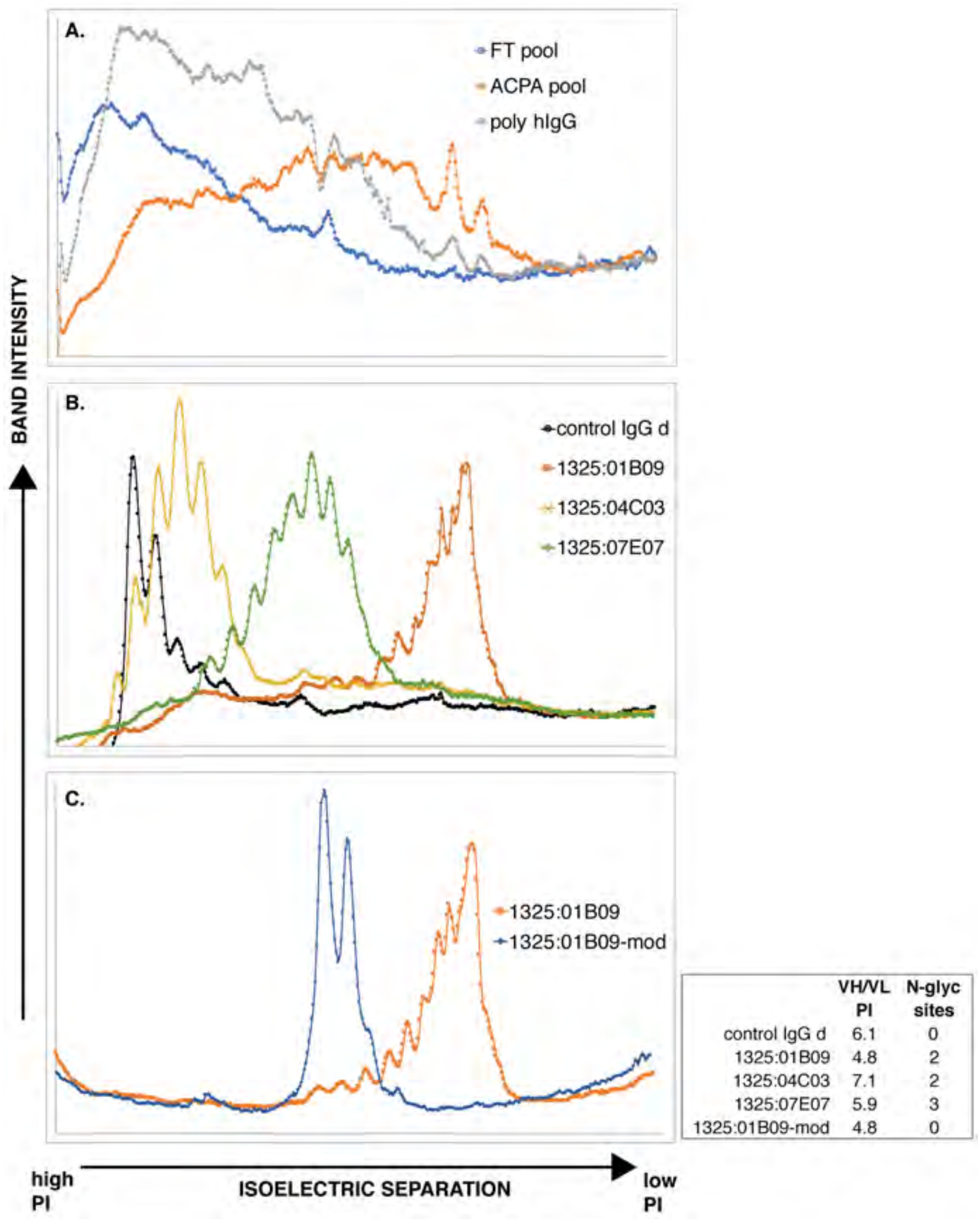

## Supporting Information Figure 9. Quantification of IgG isoforms after isoelectric focusing of ACPA

Lane profile of protein band intensity after isoelectric focusing (IEF) were quantified using ImageJ software and visualized in the graphs**. A.** Polyclonal serum affinity purified anti-CCP2 (ACPA pool) compared to the flow through IgG from the CCP column (FT pool) and purified polyclonal human IgG (Jackson ImmunoResearch). **B.** Comparing control IgG (plasma cell derived 1276:06D06 with theoretical VH/VL) with three different plasma cell derived ACPA IgG1 with different theoretical PI but all containing Fab glycosylations. **C.** Comparing the ACPA clone 1325:01B09 with a genetically modified version, 1325:01B09-mod, with the same theoretical PI but without any Fab N-glycosylation sites. Theoretical VH/VL PI and Fab N-glyc sites are listed in the box. The visualized pH range was approximately 10.7 to 4.2. All mAbs were expressed in parallel in the same system and purity was confirmed with SDS-PAGE. IEF was achieved using Novex pH 3-10 gels following the manufacturer's instructions (Thermo Fisher Scientific) and separating 10 $\mu$g polyclonal IgG or 5 $\mu$g monoclonal IgG.

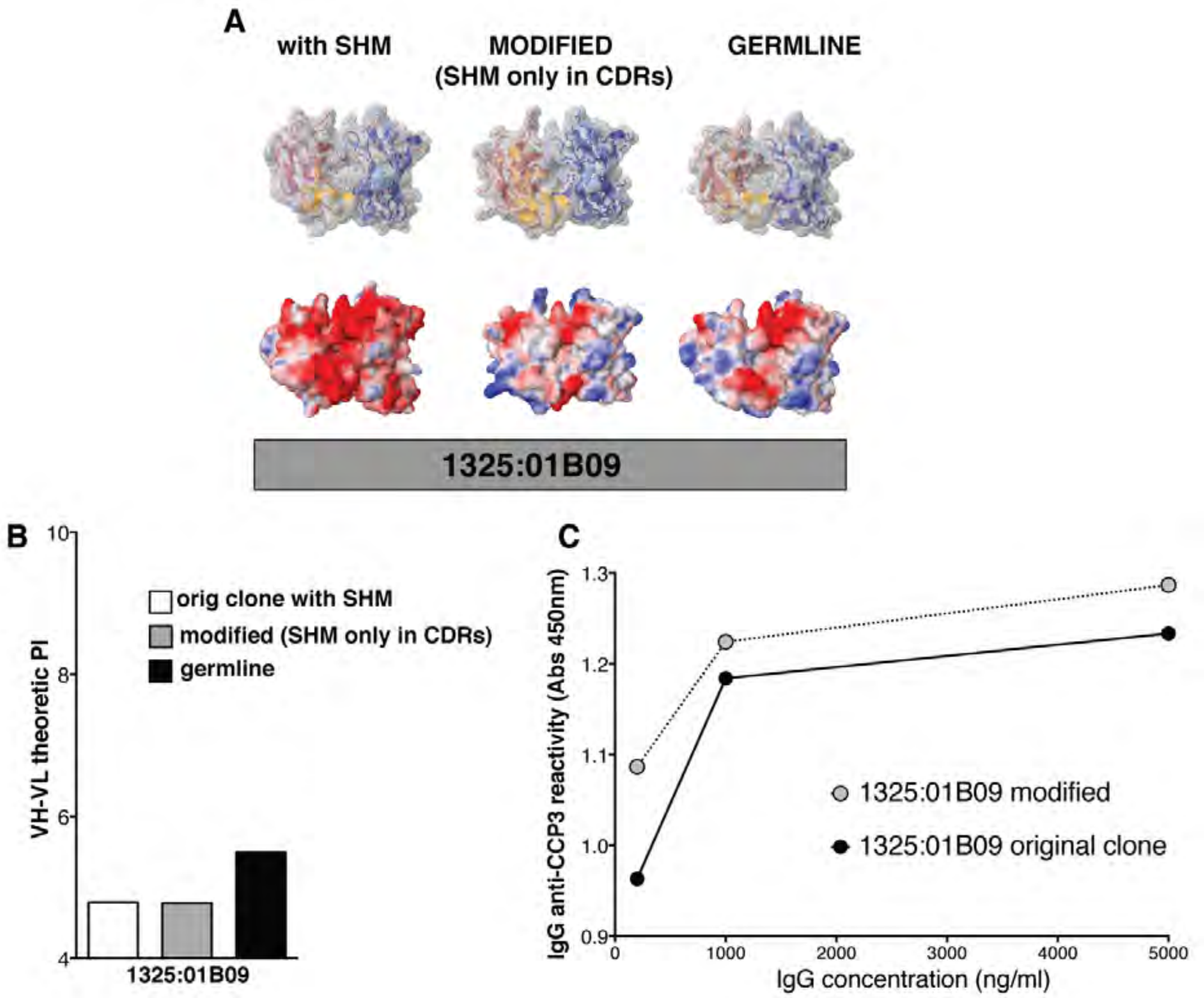

**Supporting Information Figure 10. Use of a modified ACPA variant with no N-glyc site**

A genetically modified version of the ACPA clone 1325:01B09 was designed. In this clone, somatic hypermutations in the frame work region were converted back to the closest germline sequence. This included the sequences encoding for the residues in the N-glyc sites in VH FRW1 and VL FRW1. Structure prediction (**A**) reveal some differences in surface charge but this did not change the theoretic PI compared to the original fully SHM clone (**B**). Binding to citrullinated antigens was unaffected by the changes, as demonstrated by CCP3 ELISA (**C**). Homology-based structural surface predictions were generated using PIGS web server and the best H and L chain method [7] and molecular electrostatic potential (MEP) calculations were calculated with the PDB2PQR server [8]. Structures were visualized by Jmol showing also electrostatic potential (MEP) surfaces. The RWB color (scale -.5, .5) shows positively charged surfaces as blue and negatively charged as red. The VH-VL are shown from a “top” view looking into the potential binding site.

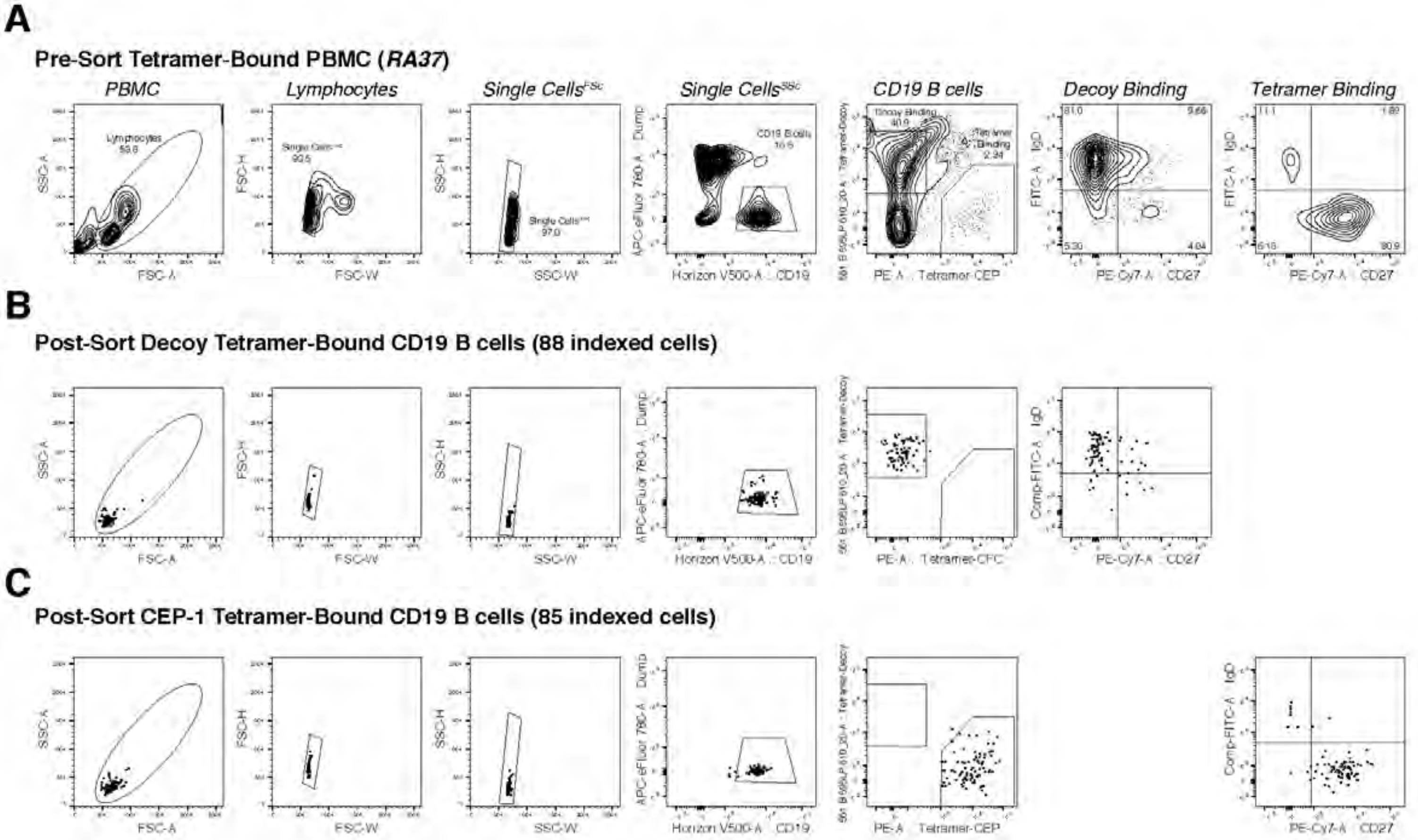

**Supporting Information Figure 11. Gating strategy for flow cytometry sorting of RA blood memory derived ACPA positive clones by antigen tetramer staining**

**A**. Pre-sort analysis of PBMC from RA patient RA37 using citrullinated $\alpha$-enolase (CEP-1) tetramer. **B.** Post-sort analysis of decoy tetramer-bound CD19 B cells (a total of 88 cells) examined in each channel. **C.** Post-sort analysis of CEP-1 tetramer-bound B cells (a total of 85 cells) examined in each channel.

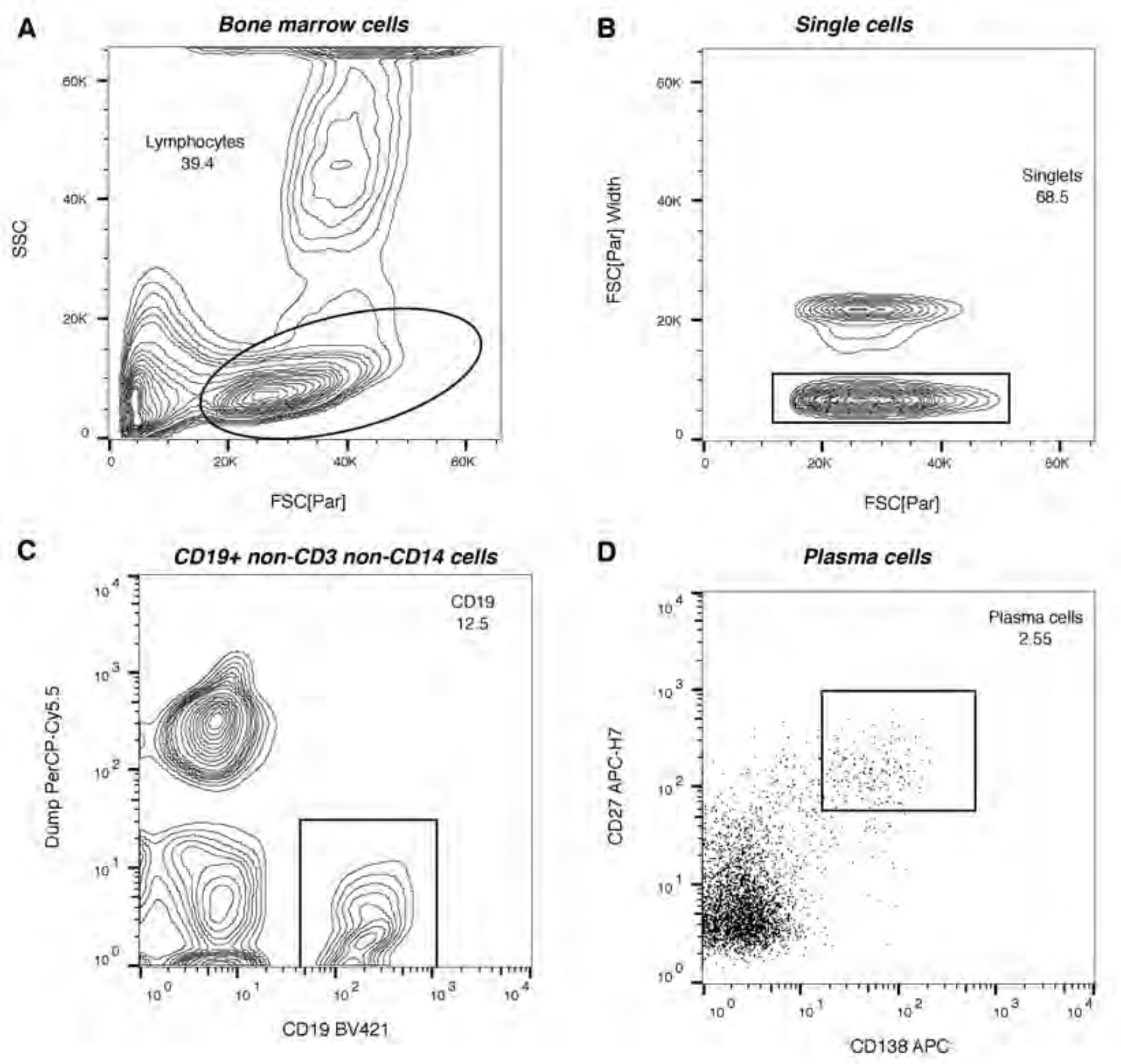

**Supporting Information Figure 12. Gating strategy for flow cytometry sorting of bone marrow plasma cells from RA patients**

Plasma cells were sorted from bone marrow mononuclear cells by gating on lymphocytes (**A**), single cells (**B**), CD19+ CD3- CD14- B cell (**C**) and CD138+ CD27+ plasma cells (**D**).

## SUPPORTING INFORMATION REFERENCES